\documentclass[fleqn,usenatbib]{mnras}

\usepackage{newtxtext,newtxmath}

\usepackage[T1]{fontenc}

\DeclareRobustCommand{\VAN}[3]{#2}
\let\VANthebibliography\thebibliography
\def\thebibliography{\DeclareRobustCommand{\VAN}[3]{##3}\VANthebibliography}

\usepackage{amsmath}
\usepackage{graphicx}
\usepackage{subfigure}
\usepackage{afterpage}
\usepackage{upgreek}

\title[Urca Production for XRBs]{Urca Nuclide Production in Type-I X-ray Bursts and Implications for Nuclear Physics Studies}

\author[Merz \& Meisel]{
Grant Merz,\thanks{gm240915@ohio.edu}
Zach Meisel\thanks{meisel@ohio.edu; Affiliated with the Joint Institute for Nuclear Astrophysics-Center for the Evolution of the Elements}
\\
Institute of Nuclear \& Particle Physics, Department of Physics \& Astronomy, Ohio University, Athens, Ohio 45701, USA\\
}

\date{Accepted XXX. Received YYY; in original form ZZZ}

\pubyear{2020}

\begin{document}
\label{firstpage}
\pagerange{\pageref{firstpage}--\pageref{lastpage}}
\maketitle

\begin{abstract}
The thermal structure of accreting neutron stars is affected by the presence of urca nuclei in the neutron star crust. Nuclear isobars harboring urca nuclides can be produced in the ashes of Type I X-ray bursts, but the details of their production have not yet been explored. Using the code {\tt MESA}, we investigate urca nuclide production in a one-dimensional model of Type I X-ray bursts using astrophysical conditions thought to resemble the source GS 1826-24.  We find that high-mass ($A\geq55$) urca nuclei are primarily produced late in the X-ray burst, during hydrogen-burning freeze-out that corresponds to the tail of the burst light curve. 
The $\sim0.4$--0.6~GK temperature relevant for nucleosynthesis of these urca nuclides is much lower than the $\sim1$~GK temperature most relevant for X-ray burst light curve impacts by nuclear reaction rates involving high-mass nuclides. The latter temperature is  often assumed for nuclear physics studies. Therefore, our findings alter the excitation energy range of interest in compound nuclei for nuclear physics studies of urca nuclide production. We demonstrate that for some cases this will need to be considered in planning for nuclear physics experiments. Additionally, we show that the lower temperature range for urca nuclide production explains why variations of some nuclear reaction rates in model calculations impacts the burst light curve but not local features of the burst ashes.
\end{abstract}



\section{Introduction} \label{sec:intro}

Type I X-ray bursts are thermonuclear explosions that occur in the envelopes of accreting neutron stars. 
Since their first observation in 1974~\citep{Grind75,Grind75b}, X-ray bursts have proved to be useful probes of accreting neutron stars thanks to their frequent recurrence~\citep{Guve13}.  Consequently, X-ray bursts aid in understanding the physics of extremely dense matter, often by way of model-observation comparisons~\citep{Stei10,Zamf12,Meis19,Good19,John20}.  However, to minimize uncertainties in model-observation comparisons for X-ray bursts and other accreting neutron star observables, the thermal structure of the neutron star outer layers should be accurately known~\citep{Meis18}.  

Recently, it was discovered that urca cooling, a neutrino-emission process long thought to operate in white dwarf stars~\citep{Tsur70,Schw17}, can significantly impact the temperature of the accreted neutron star envelope~\citep{Scha14,Deib15,Deib16,Meis17}.  Urca cooling happens when neutrinos are emitted from a cycle of repeated $e^-$-captures and $\beta^-$-decays between a pair of nuclei, creating a local heat sink~\citep{Gamo40,Gamo41}. For $e^{-}$-capture at finite temperature $T$ to a final state in a daughter nucleus within $k_{\rm B}T$ of the ground state, where $k_{\rm B}$ is the Boltzmann constant, there is phase space available for $\beta^-$-decay to occur despite the electron degeneracy of the environment. For such circumstances, and when the subsequent $e^{-}$-capture is not allowed, significant urca cooling may be possible. These conditions are only expected for nuclei with odd nuclear mass number $A$~\citep{Meis15}.

The neutrino luminosity from urca cooling is~\citep{Tsur70,Deib16}
\begin{equation}
L_{\nu} \approx L_{34}\times10^{34}{\rm{erg\,s}}{}^{-1}X(A)T_{9}^{5}\left(\frac{g_{14}}{2}\right)^{-1}R_{10}^{2} \ ,
\label{eqn:Lnu}
\end{equation}
where $X(A)$ is the mass-fraction of the $e^{-}$-capture parent nucleus in the composition and $T_{9}$ is the temperature of the urca shell in units of $10^{9} \, \mathrm{K}$. $R_{10}\equiv R/(10~\rm{km})$, where $R$ is the radius of the urca shell from the neutron star center, and $g_{14}\equiv g/(10^{14}~\rm{cm}\,\rm{s}^{-2})$, where $g = (GM/R^2)(1-2GM/Rc^2)^{-1/2}$ is the surface gravity of the neutron star with mass $M$, $c$ is the speed of light in vacuum, and $G$ is the gravitational constant.
$L_{34}(Z,A)$ is the intrinsic cooling strength of an urca pair with a parent nucleus that has $Z$ protons and $N=A-Z$ neutrons, which is determined by properties of the participating urca nuclei, such as the nuclear masses and weak transition rates (see Equation~\ref{eqn:L34})~\citep{Meis18}.

$X(A)$ is set by the ashes of nuclear burning occurring in the neutron star envelope. In the ocean and outer neutron star crust, $X(A)$ from surface burning processes such as X-ray bursts are preserved since burial of ashes by subsequent accretion initially results in transmutation by $e^{-}$-capture\footnote{The majority of crust reactions modifying $X(A)$, such as $e^{-}$-capture neutron-emission reactions, density driven fusion, and possibly neutron transfers, tend to occur at deeper depths than are relevant for urca cooling~\citep{Lau18,Chug19}. While neutron transfers may burn-out some urca nuclides~\citep{Chug19}, a detailed investigation of their impact has not yet been performed and current rate estimates are rough approximations. Semi-quantitative estimates suggest that neutron transfers may actually enhance the abundance of some urca nuclides at the expense of others.}. While past numerical calculations have reported $X(A)$ from X-ray bursts that are often substantial for odd-$A$ nuclides, e.g.~\citep{Woos04,Fisk08,Jose10,Cybu16,Meis19}, the details of odd-$A$ nuclide production have not yet been described. This is potentially problematic, as nuclear reaction rate uncertainties are known to affect $X(A)$ for urca nuclides up to two orders of magnitude~\citep{Cybu16,Meis19}. Without knowing the relevant environment temperatures, nuclear physicists lack guidance required to adequately plan for studies of these reactions.

We use one-dimensional models of type-I X-ray bursts to demonstrate that urca nuclides with $A\geq55$ are produced in the hydrogen-burning freeze-out near the end of the burst. In Section~\ref{sec:mesa} we describe our X-ray burst model calculations.  Section~\ref{sec:results} presents calculation results, demonstrating the late-time urca nuclide production. Section~\ref{sec:ASTimplications} demonstrates the sensitivity of urca nuclide production to nuclear reation rate variations. Section~\ref{sec:NUCimplications} discuss the implications for nuclear physics studies. We summarize our work in Section~\ref{sec:concl}.

\section{Model calculations} \label{sec:mesa}

The code {\tt MESA}~\citep{Paxt11,Paxt13,Paxt15} version 9793 was used for one-dimensional model calculations of Type-I X-ray bursts, following the approach described in~\citet{Meis18b,Meis19}. The most pertinent details are repeated here. We followed burst nucleosynthesis within $\sim$1000 zones constructed to resemble an accreting neutron star envelope. General relativistic effects were accounted for using a post-Newtonian modification to the local gravity~\citep{Paxt11}. Convection was approximated using the time-dependent mixing length theory of~\citet{Heny65}. Adaptive time and spatial resolution were employed according the {\tt MESA} controls {\tt varcontrol\_target=1d-3} and {\tt mesh\_delta\_coeff=1.0}~\citep{Paxt13}. The inner boundary of the $\sim$0.01~km thick envelope is treated as if it is on top of a neutron star with $M=1.4~M_{\odot}$ and $R=11.2$~km. Nuclear reaction rates from REACLIB~\citep{Cybu10} version 2.2 were used within the 304~isotope network developed by \citet{Fisk08} for analyses of the rapid proton-capture ($rp$)-process.

Calculations employed the baseline accretion conditions of \citet{Meis19}, which were found by \citet{Meis18b}\footnote{See also fig. 6 of \citet{Adsl20} for a light curve model-observation comparison.} to best reproduce observed features for the textbook X-ray burst source GS 1826-24~\citep{Bild00,Gall08}. These conditions include accretion rate $\dot{M} = 0.17~\dot{M_{\rm E}}$ with the Eddington accretion rate $\dot{M_{\rm E}}=1.75\times10^{-8}~M_\odot$\,yr$^{-1}$ from \citet{Schatz99}, accretion-based heating at the base
of the envelope $Q_{b}$ = 0.1~MeV\,u$^{-1}$, accreted hydrogen mass fraction $X_{\rm H}=0.70$, accreted metal mass fraction $Z=0.02$ with the solar metal distribution of \citet{Grev98}, and accreted helium mass fraction $Y=1-X_{\rm H}-Z$. To mitigate the effects of compositional inertia~\citep{Woos04}, we calculated a sequence of twenty X-ray bursts, where the results of the twentieth burst are shown here.

\section{Results and Discussion}
\label{sec:results}

\begin{figure*}
  \centering
  \subfigure[$A=29$]{\label{fig:a29}\includegraphics[width=2.8in]{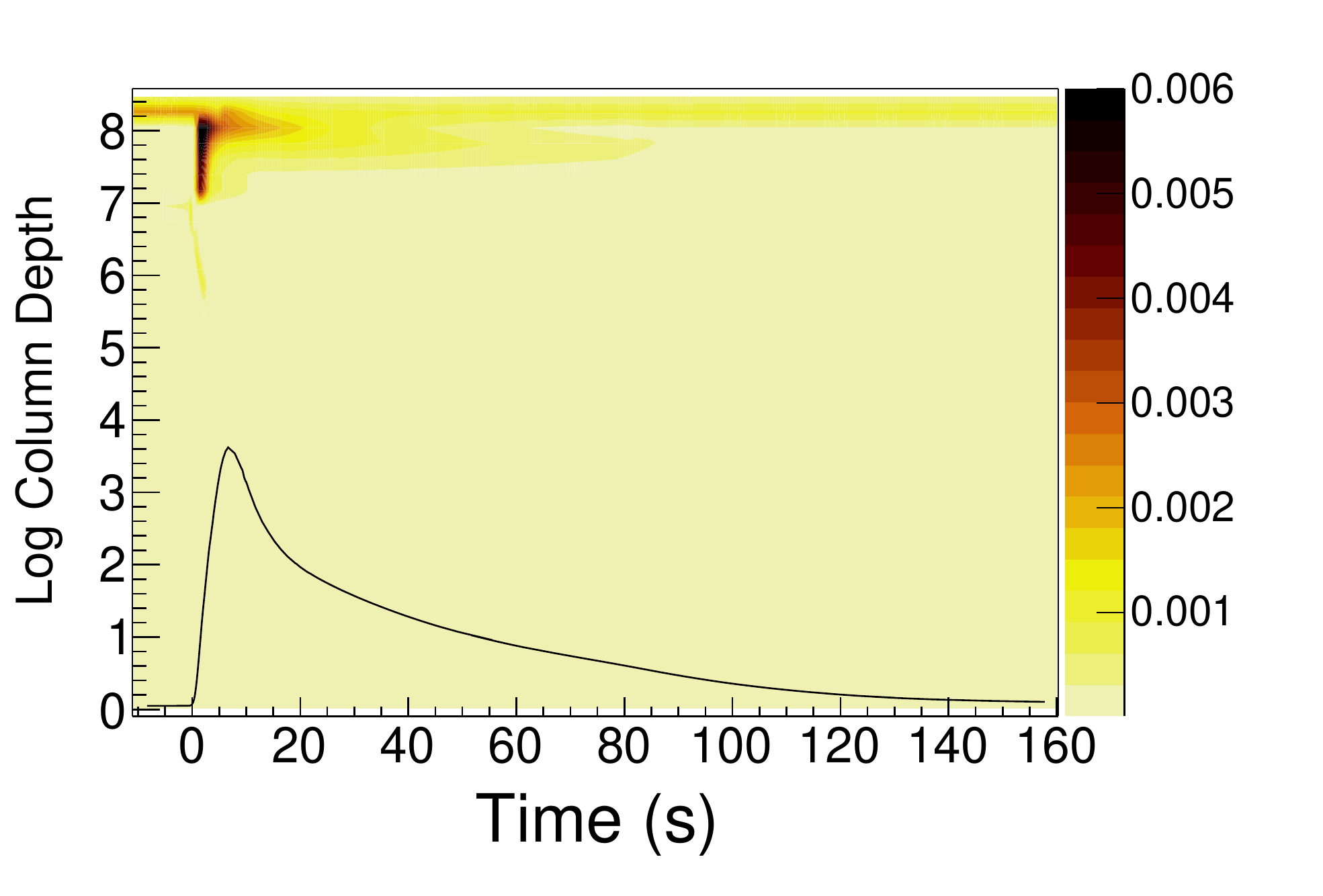}}
  \qquad
  \subfigure[$A=31$]{\label{fig:a31}\includegraphics[width=2.8in]{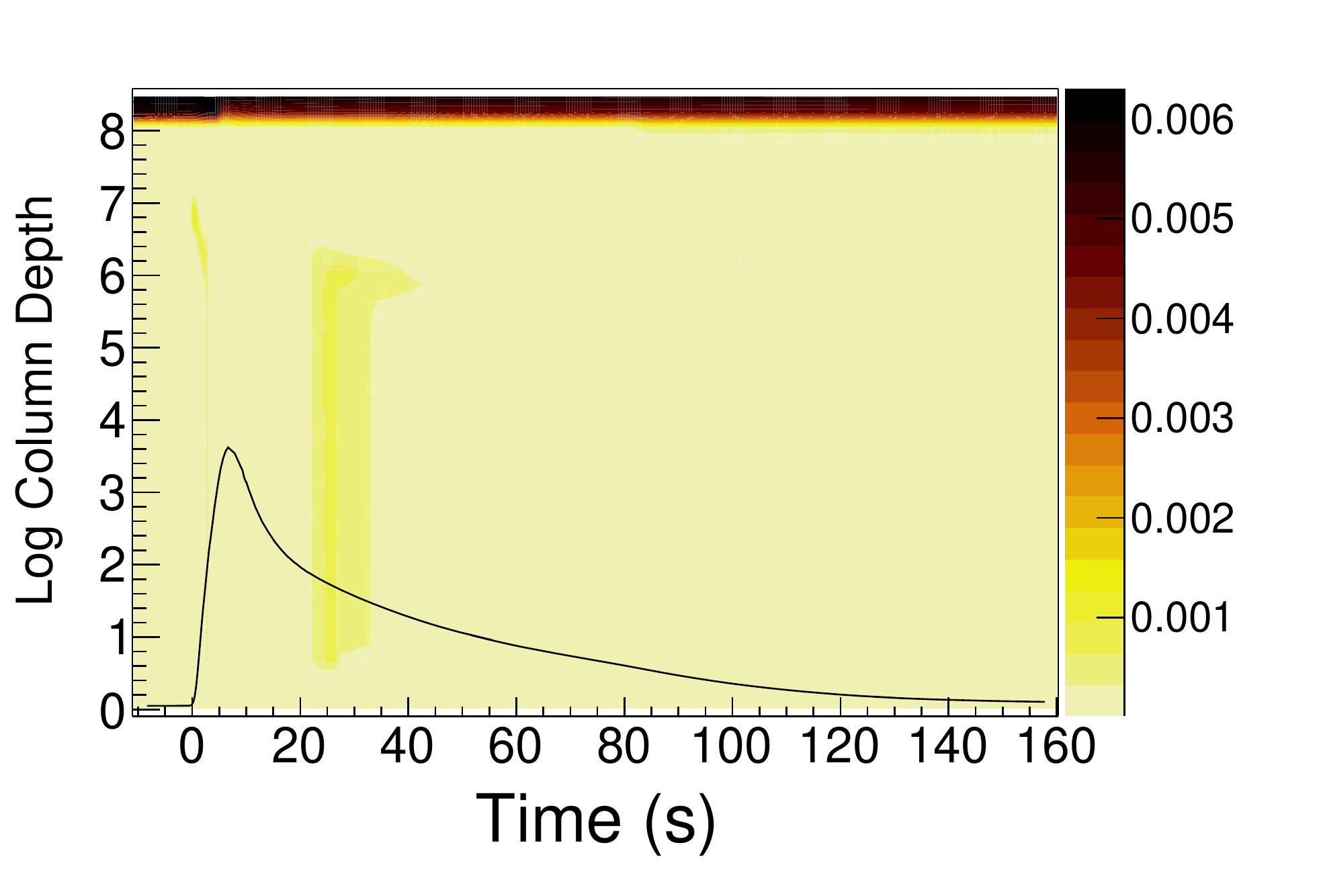}}
  \medskip
  \subfigure[$A=33$]{\label{fig:a33}\includegraphics[width=2.8in]{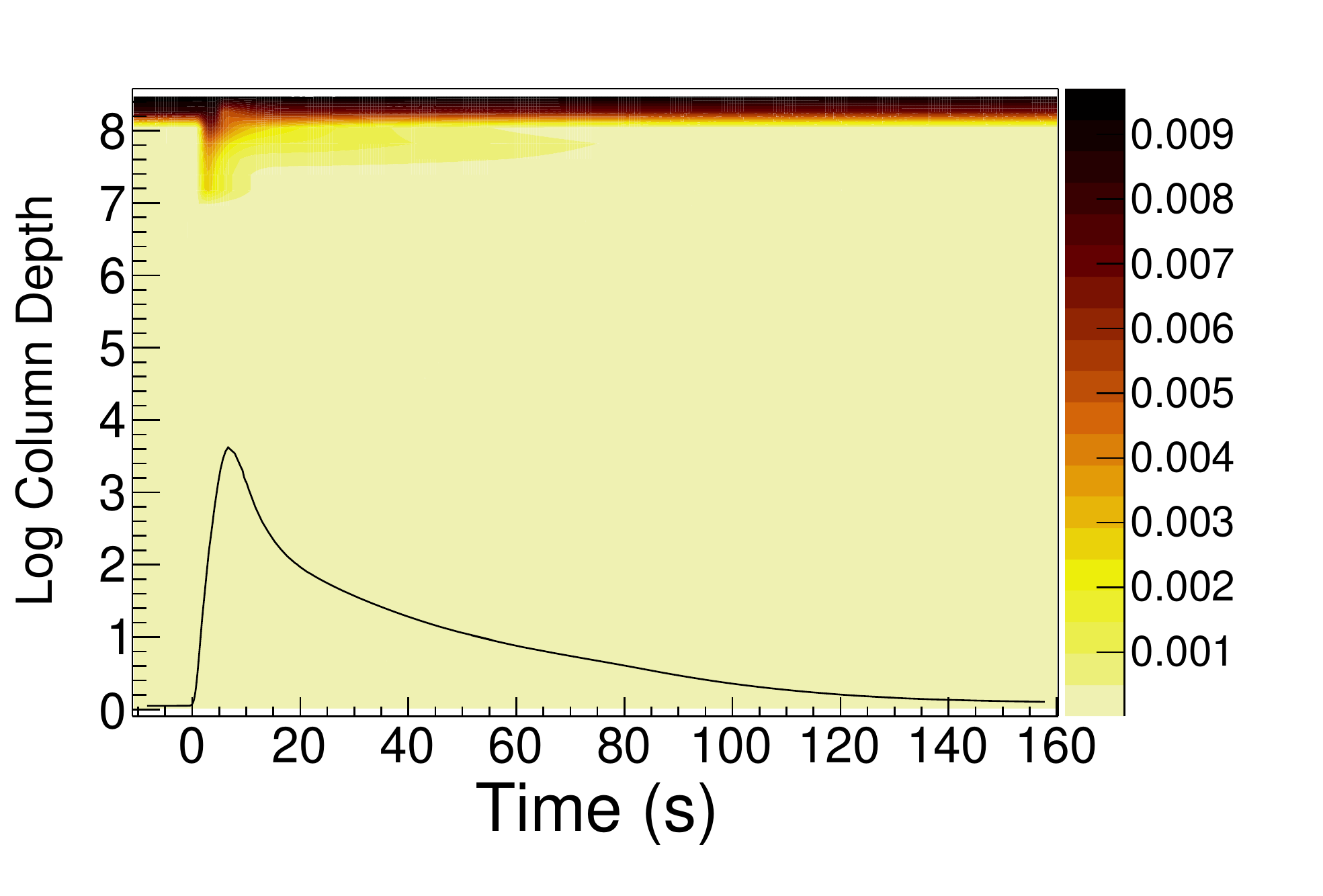}}
 \qquad
  \subfigure[$A=55$]{\label{fig:a55}\includegraphics[width=2.8in]{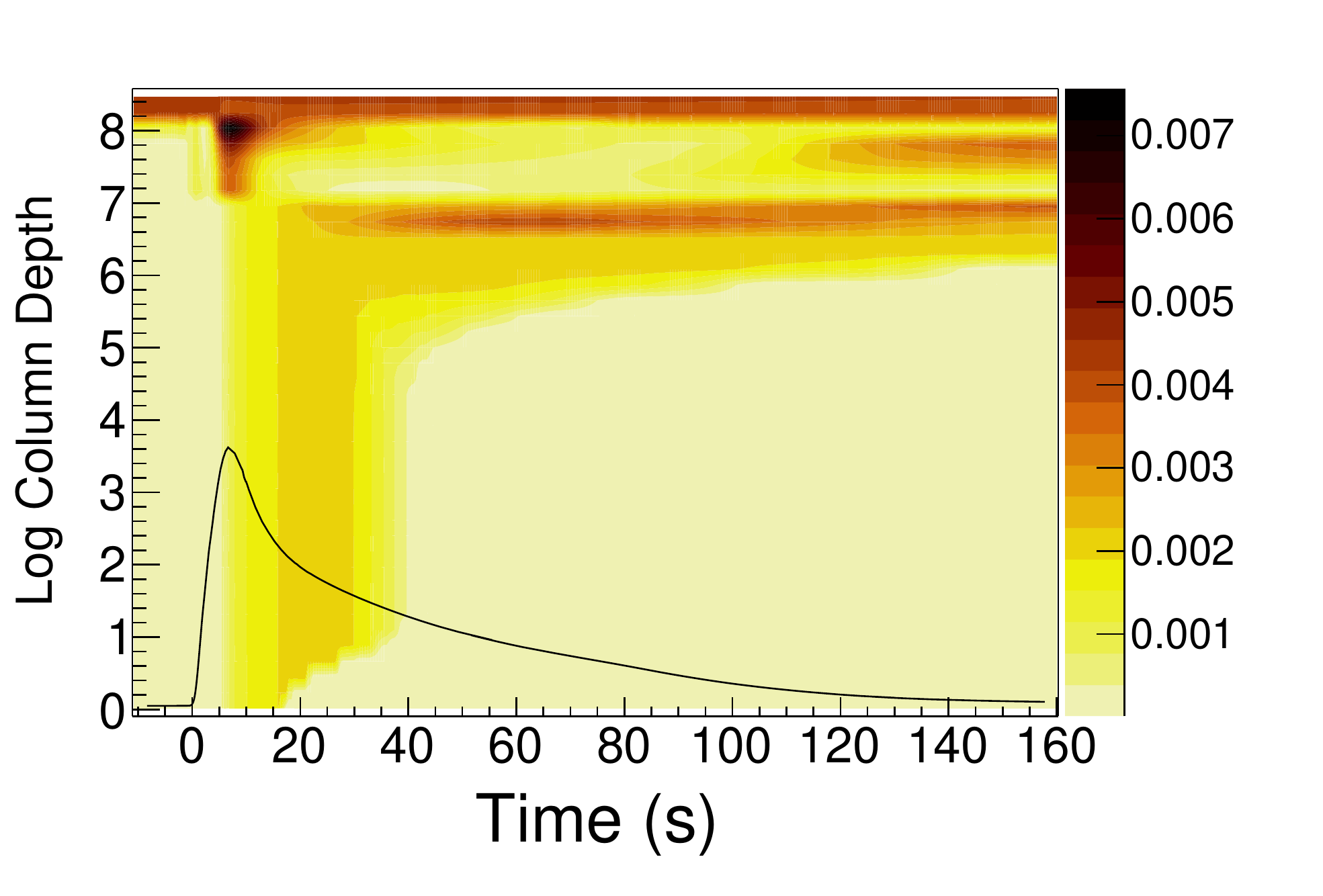}}
   \medskip
  \subfigure[$A=57$]{\label{fig:a57}\includegraphics[width=2.8in]{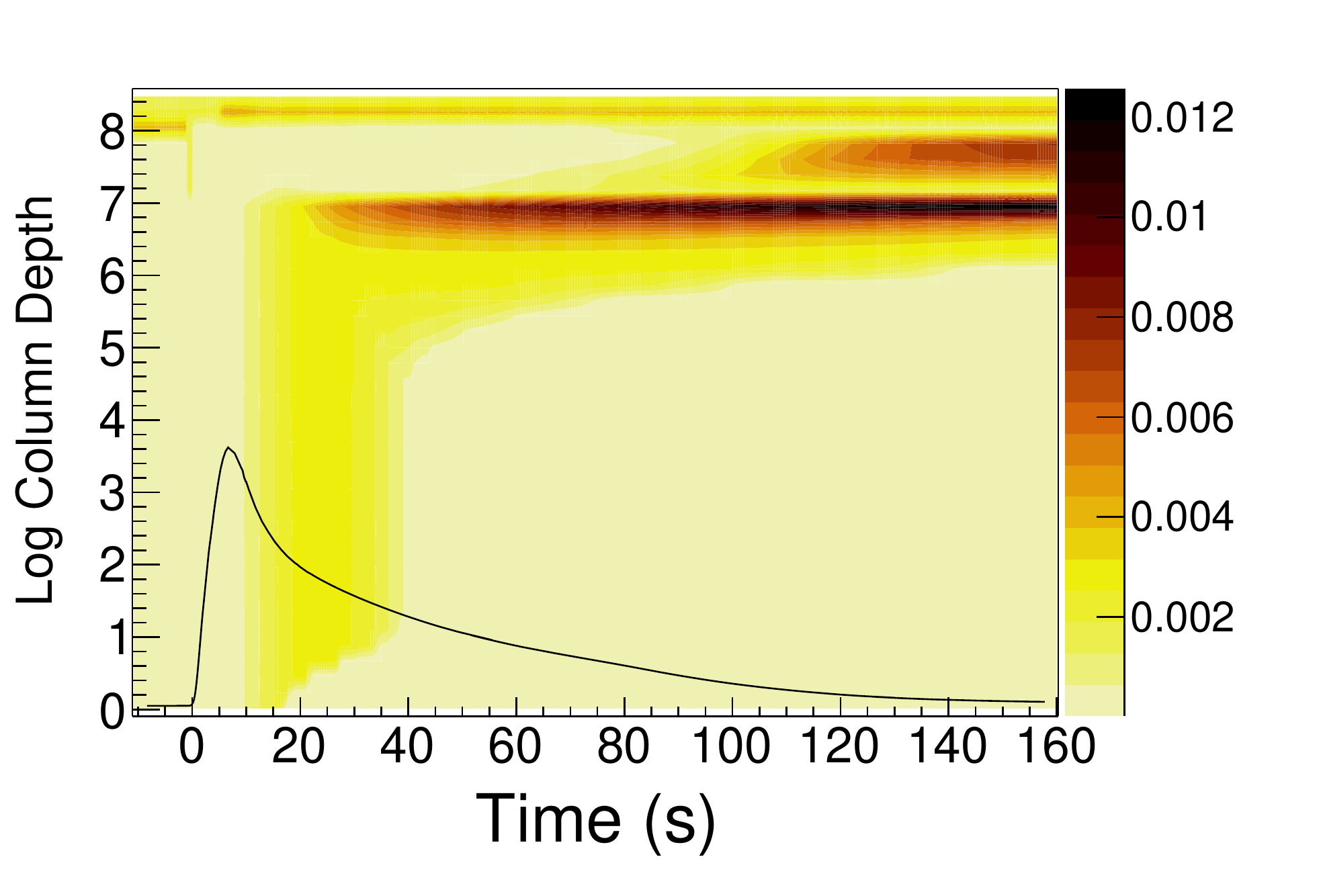}}
  \qquad
  \subfigure[$A=61$]{\label{fig:a61}\includegraphics[width=2.8in]{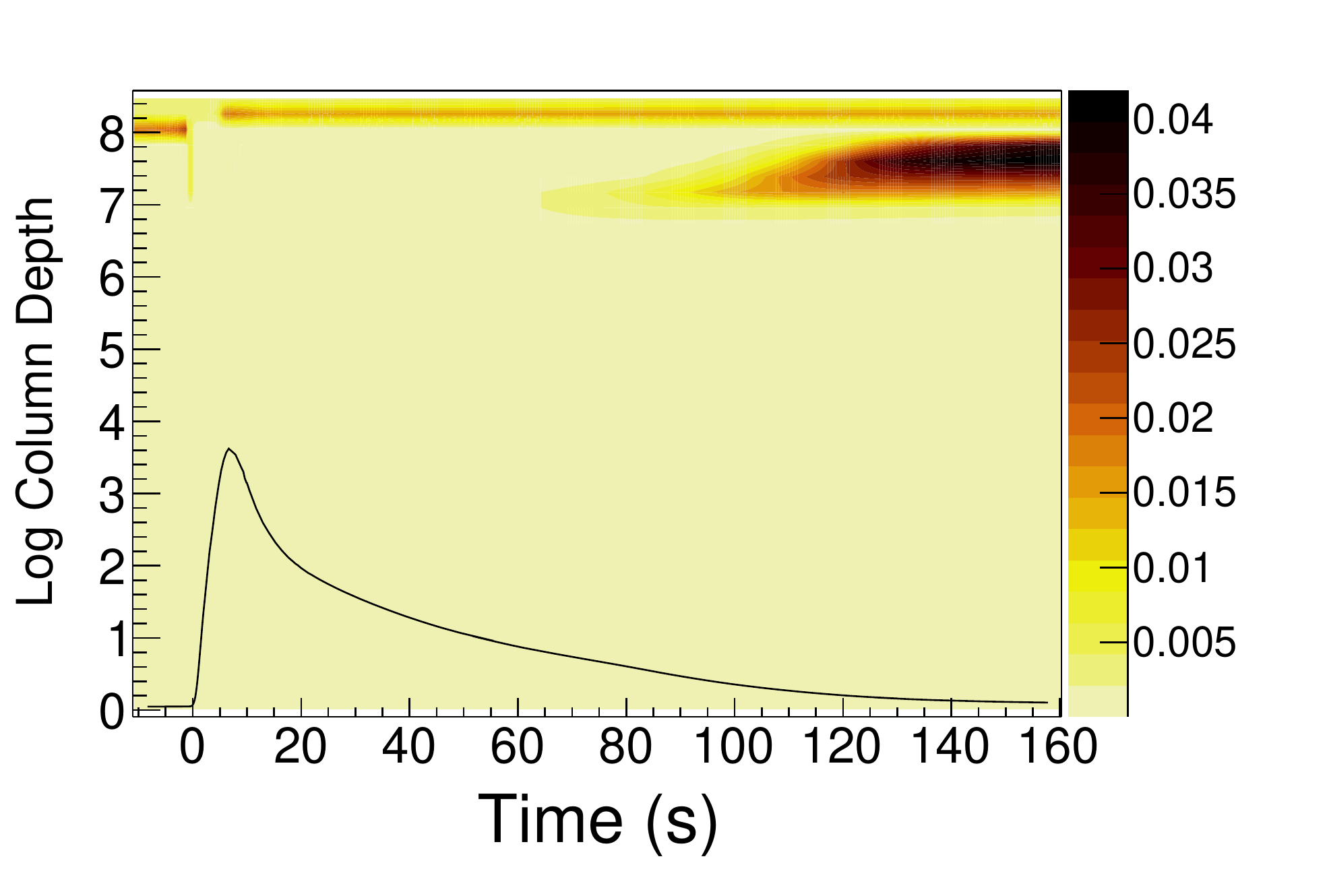}}
   \medskip
  \subfigure[$A=63$]{\label{fig:a63}\includegraphics[width=2.8in]{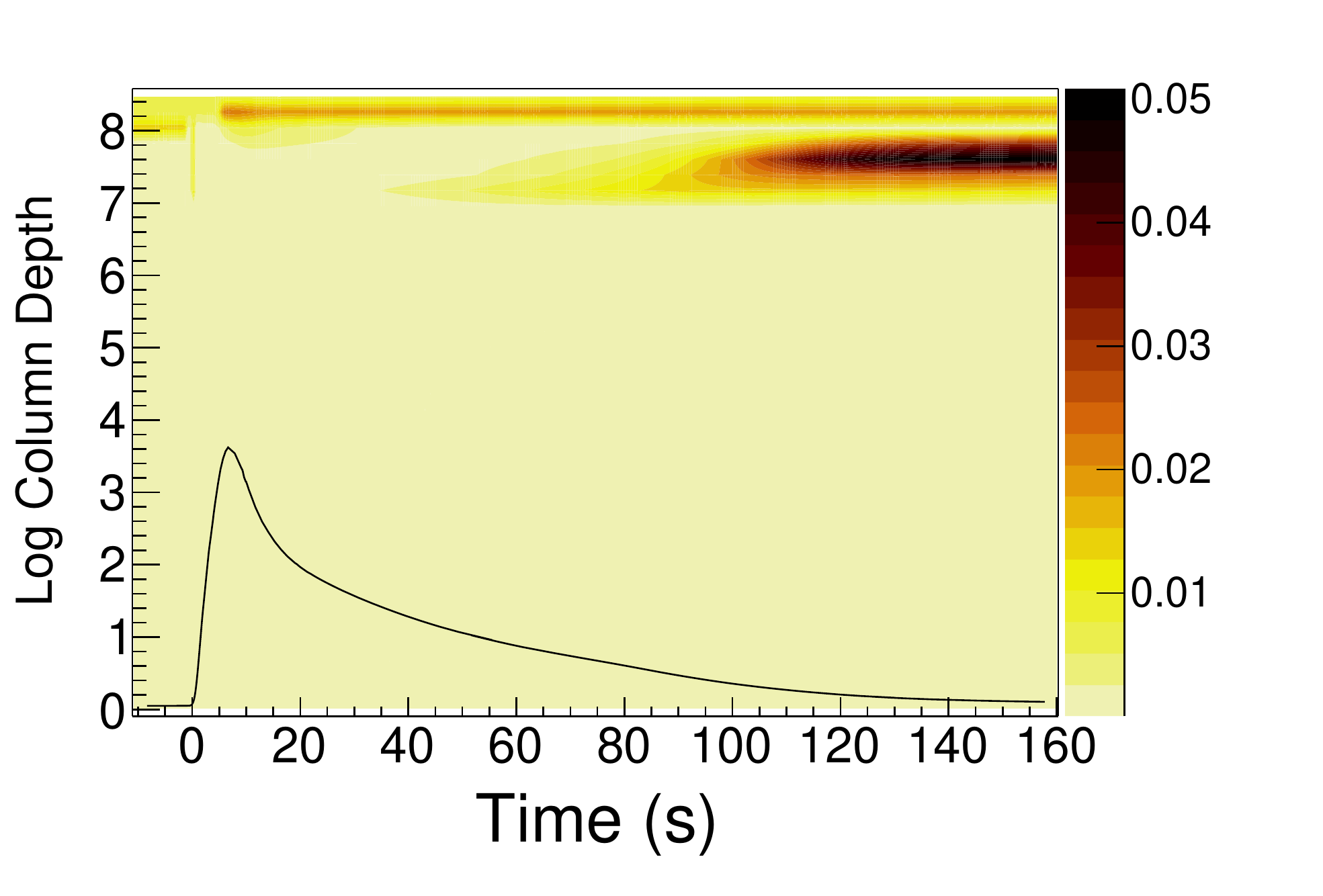}}
    \qquad
  \subfigure[$A=65$]{\label{fig:a65}\includegraphics[width=2.8in]{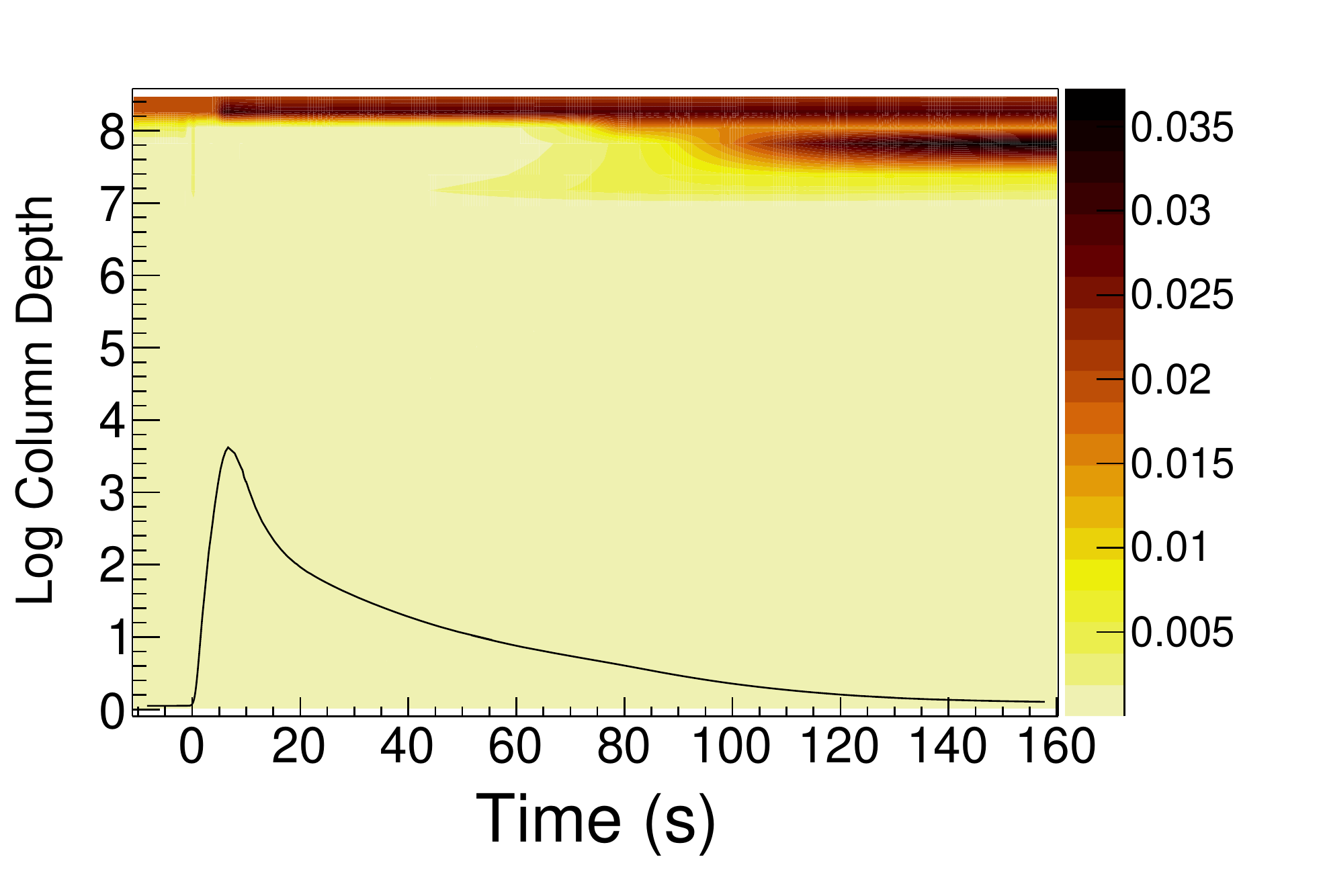}}
 \caption{Depth profile of $X(A)$ over the burst evolution, where the light curve is shown for reference.}
 \label{fig:XA}
\end{figure*}

\begin{figure}
    \centering
     \subfigure[$A=59$]{\label{fig:a59}\includegraphics[width=0.5\textwidth]{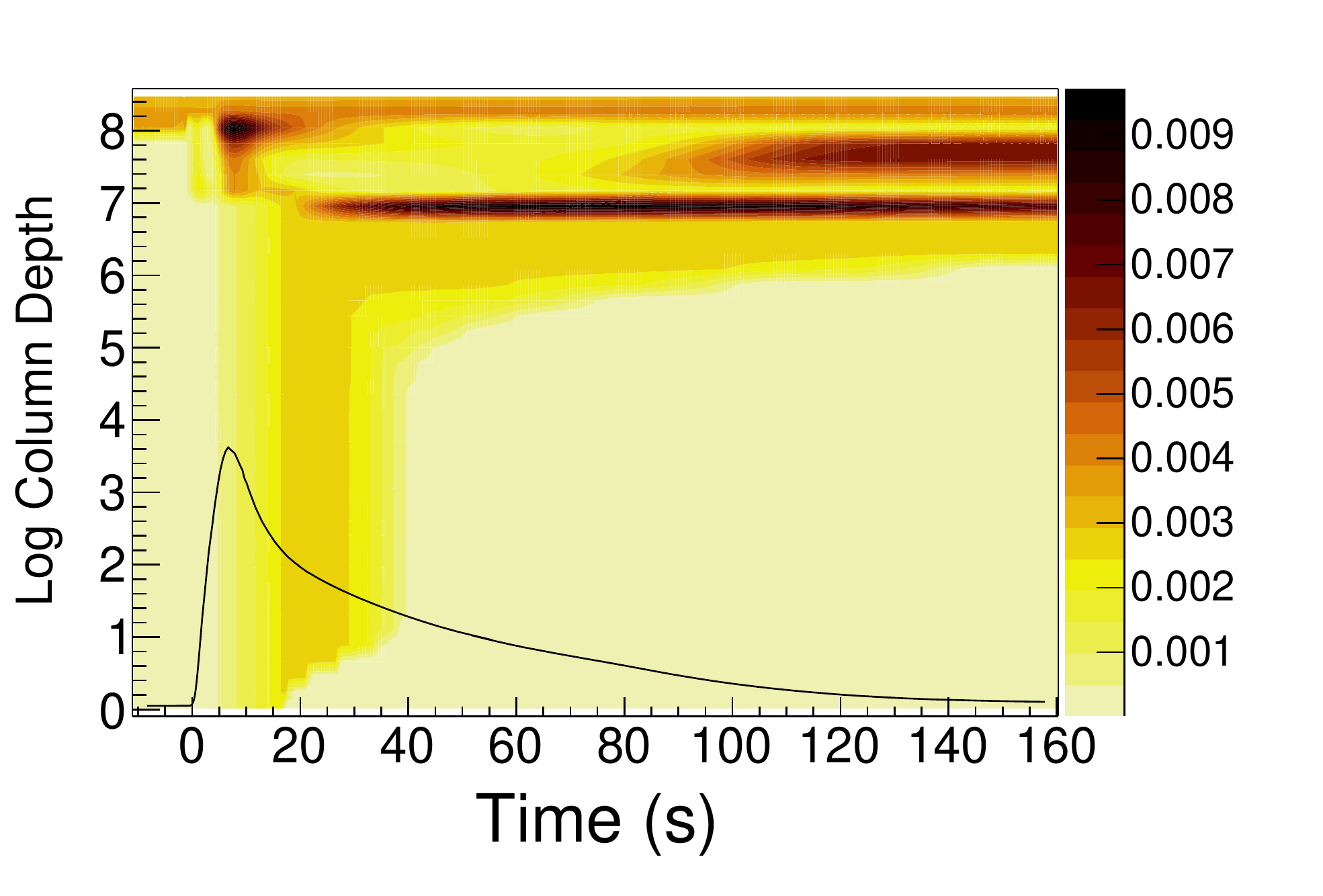}}
  \medskip
  \subfigure[Hydrogen]{\label{fig:aH}\includegraphics[width=0.5\textwidth]{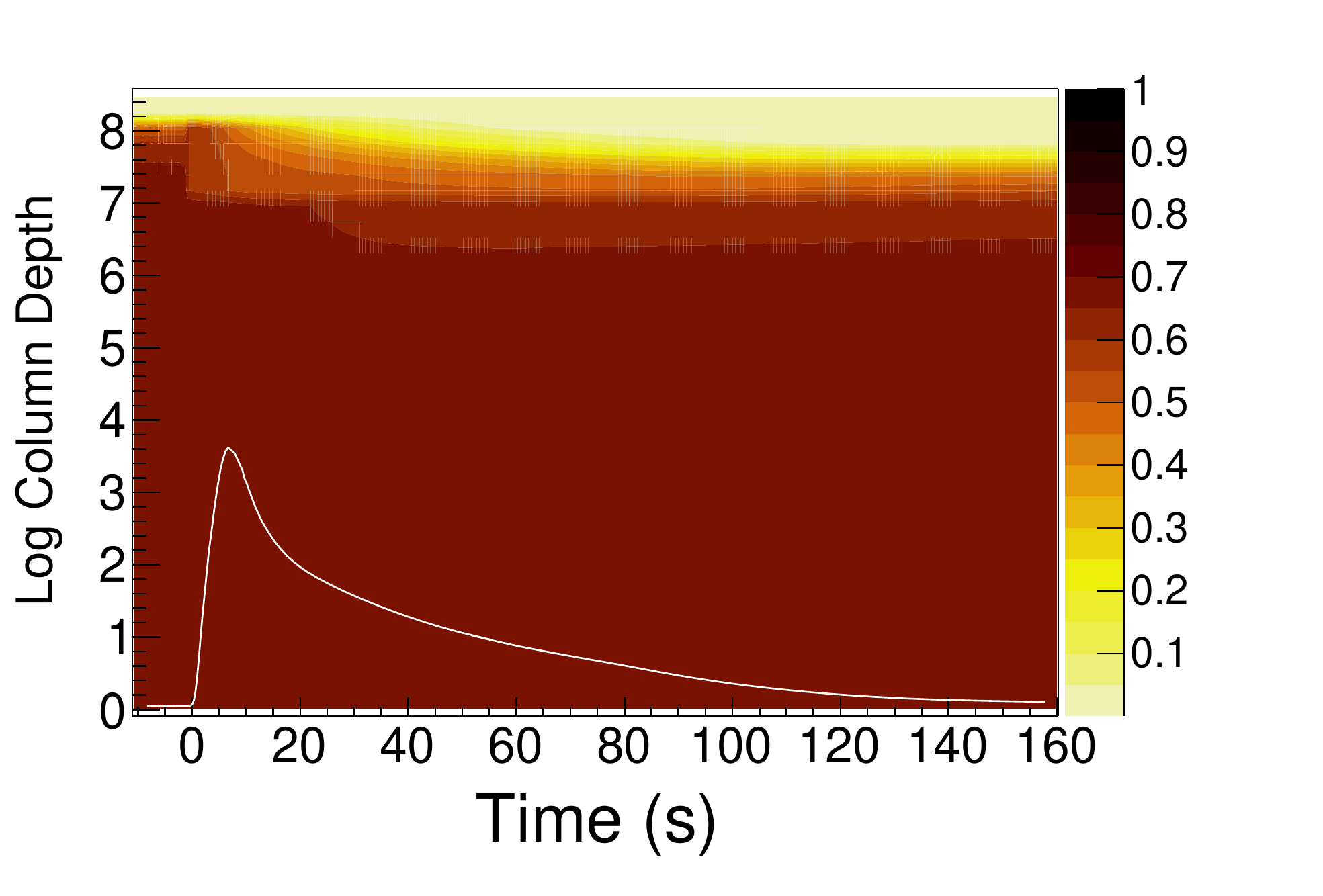}}
    \caption{Same as Figure~\ref{fig:XA}, for $A=59$ and hydrogen.}
    \label{fig:XH}
\end{figure}

Figs.~\ref{fig:XA} and~\ref{fig:XH} show $X(A)$ for hydrogen and most suspected urca nuclides~\citep{Scha14,Meis17} throughout the neutron star envelope over the X-ray burst evolution, where the column depth $y\equiv \int_{r}^{\infty} \rho\,{\rm d}r'$, $\rho$ is the mass density, and $r$ is the radial coordinate. Of urca nuclides listed in \citet{Scha14} and \citet{Meis17}, we only include odd-$A$ nuclides, as it is unlikely that even-$A$ nuclides can satisfy the conditions for urca cycling~\citep{Meis15}. Additionally, $A=103$ and 105 are not investigated as they are not produced in appreciable quantities in the model calculations presented here (see~\citet{Meis19}) and $A=61$ is included based on the results of \citet{Ong18}.

\begin{figure*}
    \centering
 \subfigure[]{\label{fig:sparknocut}\includegraphics[width=0.41\textwidth]{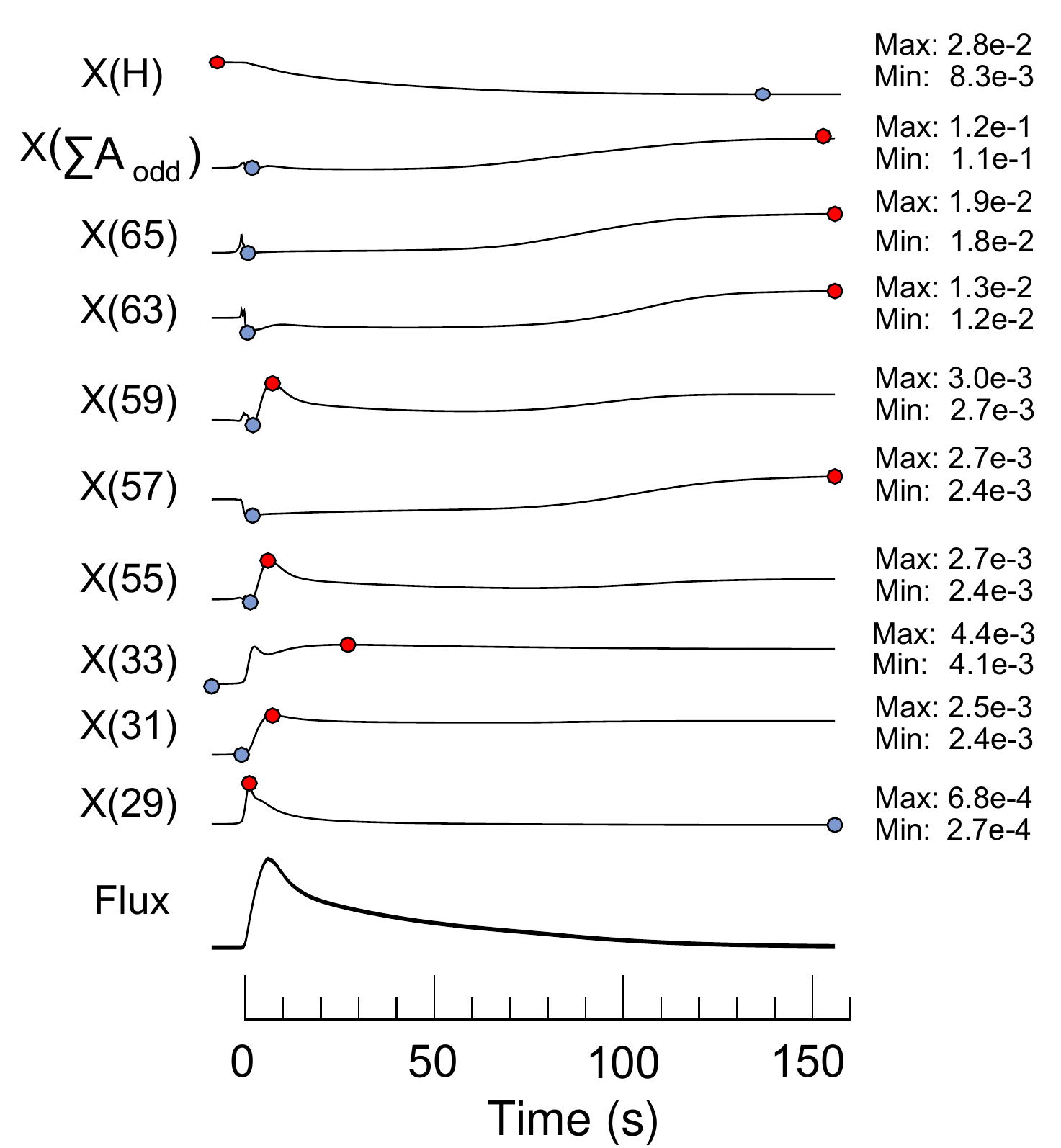}}
 \medskip
  \subfigure[]{\label{fig:spark}\includegraphics[width=0.38\textwidth]{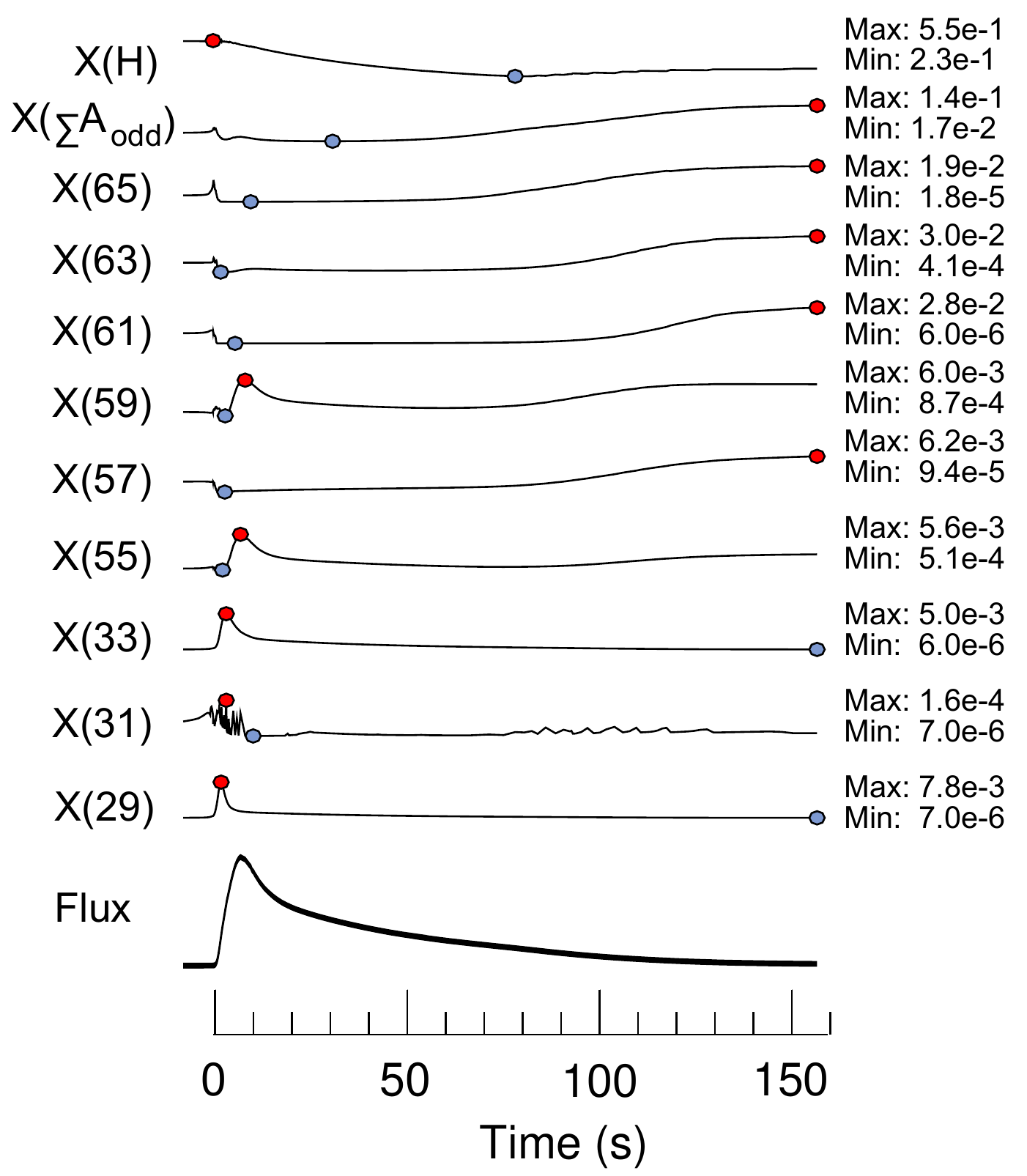}}
    \caption{Sparklines~\citep{tuft01} comparing the time evolution of select $X(A)$ for a weighted average over (a) the entire neutron star envelope above the iron substrate and (b) only the region with hydrogen burning at late times, where the X-ray flux is shown for reference. Blue (red) circles indicate the minimum (maximum) for a quantity, with corresponding values indicated in the rightward legend. Note that the slight increase in $X(H)$ late in the burst evolution is due to replenishment from accretion.}
    \label{fig:sparks}
\end{figure*}

\begin{figure}
    \centering
    \subfigure[$\rho~{\rm[}10^{6}\,{\rm g\,cm}^{-3}{\rm]}$]{\label{fig:rho}\includegraphics[width=0.5\textwidth]{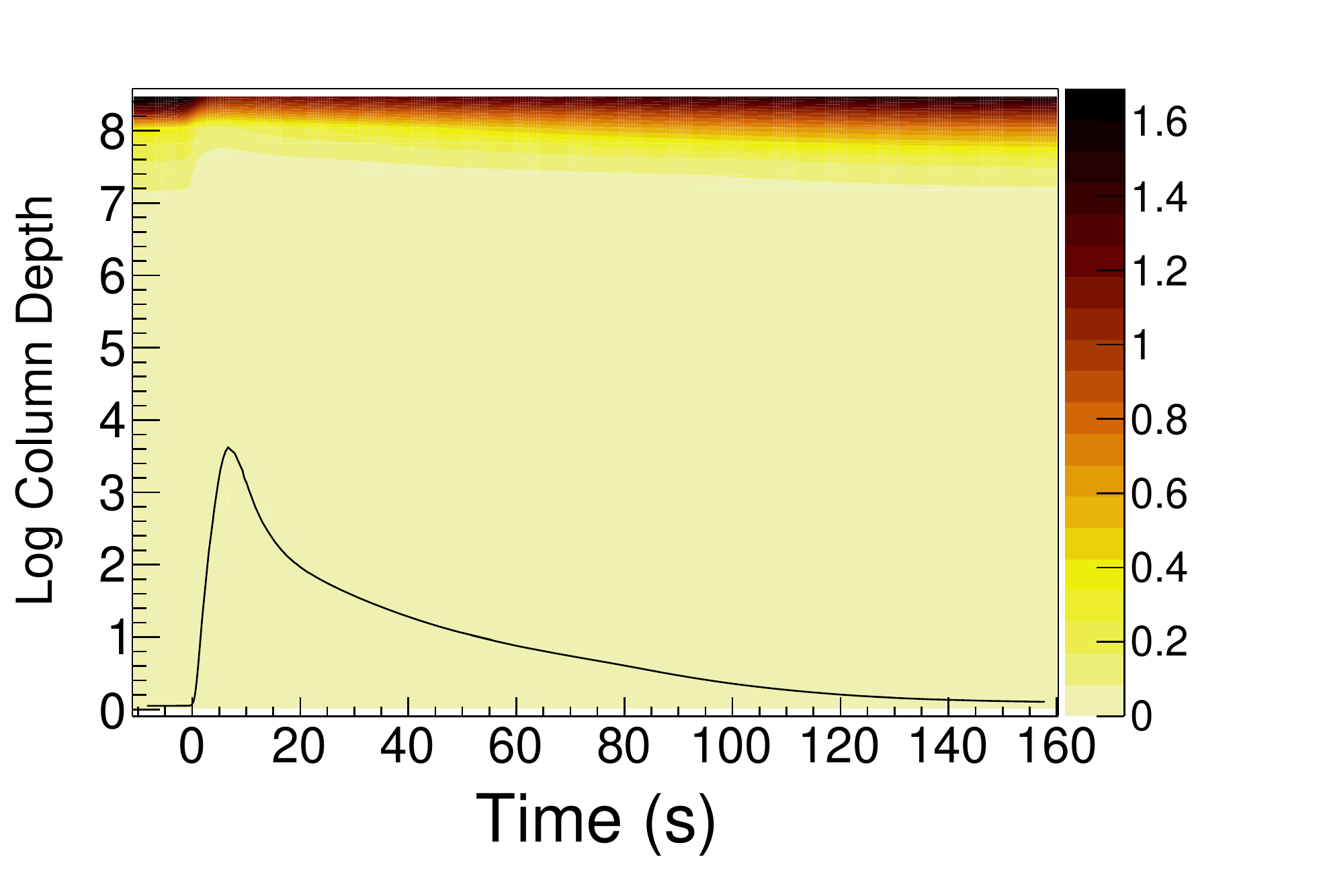}}
  \qquad
  \subfigure[$T~{\rm[GK]}$]{\label{fig:temp}\includegraphics[width=0.5\textwidth]{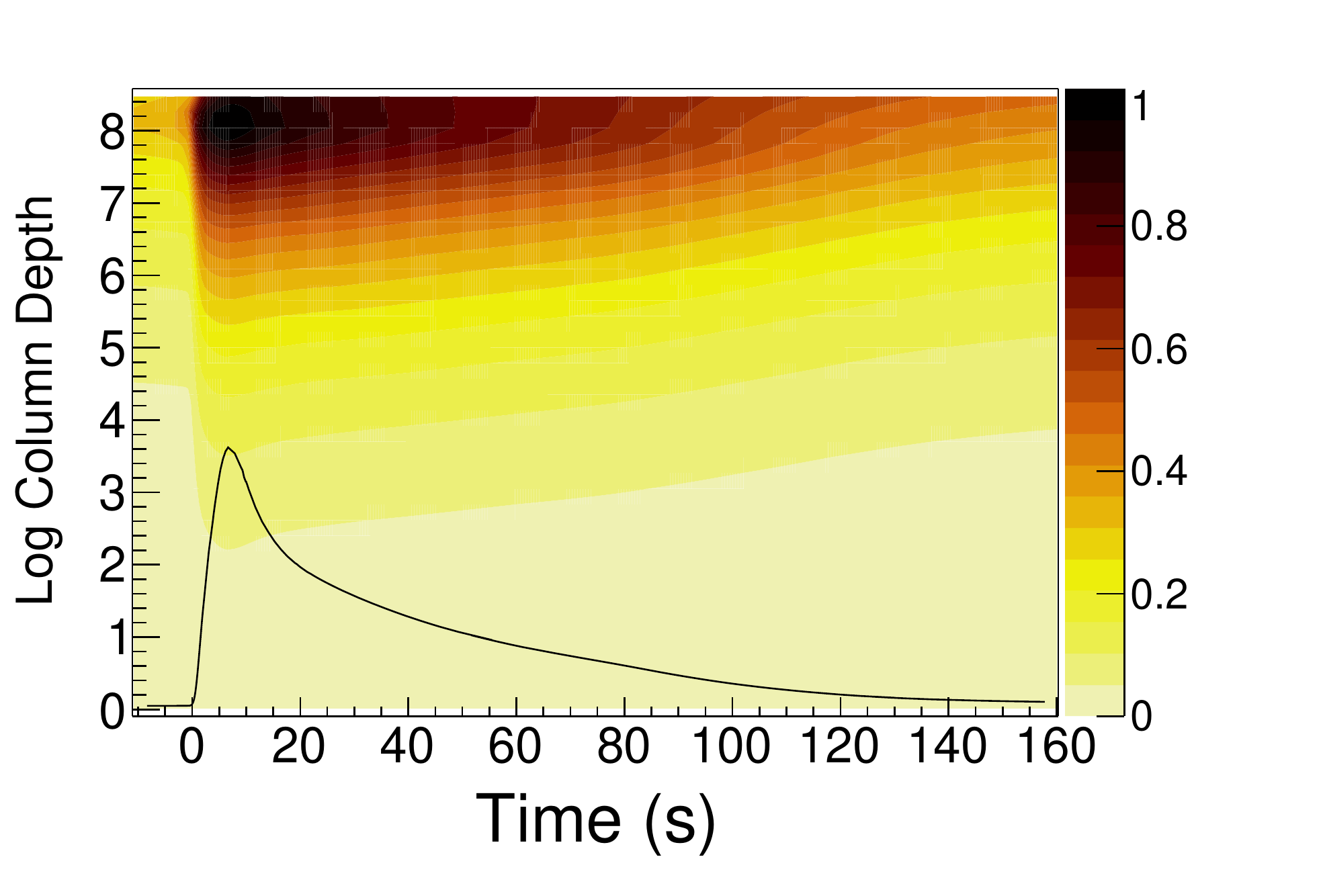}}
    \caption{Depth profile of $\rho$ and $T$ over the burst evolution, where the light curve is shown for reference.}
    \label{fig:denstemp}
\end{figure} 

\begin{table}
\centering
\caption{\label{tab:lambda} $\lambda$ for urca nuclide production ($+$) and destruction ($-$) reactions. $t=60$~s corresponds to conditions $X({\rm H})=0.4$, $T=0.6$~GK, and $\rho=10^{5}$~g\,cm$^{-3}$. For $t=120$~s, $X({\rm H})=0.1$, $T=0.4$~GK, and $\rho=10^{5}$~g\,cm$^{-3}$. Units for $N_{\rm A}\langle\sigma v\rangle$ are [cm$^{3}$\,mol$^{-1}$\,s$^{-1}$].}
\begin{tabular}{c|c|c|ccc}%
\hline\hline
$A$ & Rate & $+/-$ & $t$~[s] & $N_{\rm A}\langle\sigma v\rangle$ & $\lambda$~[s$^{-1}$] \\
\hline
 & $^{28}{\rm P}(p,\gamma)$ & + & 60 & $1.1\times10^{0}$ & $4.4\times10^{4}$ \\
\cline{4-6}
 29 &  &  & 120 & $3.8\times10^{-2}$ & $3.8\times10^{2}$ \\
\cline{2-6}
 & $^{29}{\rm P}(p,\gamma)$ & - & 60 & $1.5\times10^{-1}$ & $6.0\times10^{3}$ \\
\cline{4-6}
  &  &  & 120 & $7.1\times10^{-3}$ & $7.1\times10^{1}$ \\
\hline

 & $^{30}{\rm P}(p,\gamma)$ & + & 60 & $2.6\times10^{0}$ & $1.0\times10^{5}$ \\
\cline{4-6}
 31 &  &  & 120 & $5.9\times10^{-2}$ & $5.9\times10^{2}$ \\
\cline{2-6}
 & $^{31}{\rm S}(p,\gamma)$ & - & 60 & $2.2\times10^{-2}$ & $8.8\times10^{2}$ \\
\cline{4-6}
  &  &  & 120 & $1.4\times10^{-4}$ & $1.4\times10^{0}$ \\
\hline

 & $^{32}{\rm Cl}(p,\gamma)$ & + & 60 & $3.0\times10^{-2}$ & $1.2\times10^{3}$ \\
\cline{4-6}
 33 &  &  & 120 & $4.1\times10^{-4}$ & $4.1\times10^{0}$ \\
\cline{2-6}
 & $^{33}{\rm Cl}(p,\gamma)$ & - & 60 & $3.0\times10^{-1}$ & $1.2\times10^{4}$ \\
\cline{4-6}
  &  &  & 120 & $6.1\times10^{-3}$ & $6.1\times10^{1}$ \\
\hline

 & $^{54}{\rm Co}(p,\gamma)$ & + & 60 & $1.5\times10^{-4}$ & $6.0\times10^{0}$ \\
\cline{4-6}
 55 &  &  & 120 & $4.0\times10^{-7}$ & $4.0\times10^{-3}$ \\
\cline{2-6}
 & $^{55}{\rm Co}(p,\gamma)$ & - & 60 & $1.1\times10^{-4}$ & $4.4\times10^{0}$ \\
\cline{4-6}
  &  &  & 120 & $3.4\times10^{-7}$ & $3.4\times10^{-3}$ \\
\hline

 & $^{56}{\rm Ni}(p,\gamma)$ & + & 60 & $2.1\times10^{-5}$ & $8.4\times10^{-1}$ \\
\cline{4-6}
 57 &  &  & 120 & $7.1\times10^{-7}$ & $7.1\times10^{-3}$ \\
\cline{2-6}
 & $^{57}{\rm Ni}(p,\gamma)$ & - & 60 & $4.2\times10^{-5}$ & $1.7\times10^{0}$ \\
\cline{4-6}
  &  &  & 120 & $1.2\times10^{-7}$ & $1.2\times10^{-3}$ \\
\hline

 & $^{58}{\rm Cu}(p,\gamma)$ & + & 60 & $2.2\times10^{-4}$ & $8.8\times10^{0}$ \\
\cline{4-6}
 59 &  &  & 120 & $3.3\times10^{-7}$ & $3.3\times10^{-3}$ \\
\cline{2-6}
 & $^{59}{\rm Cu}(p,\gamma)$ & - & 60 & $2.9\times10^{-5}$ & $1.2\times10^{0}$ \\
\cline{4-6}
  &  &  & 120 & $5.9\times10^{-8}$ & $5.9\times10^{-4}$ \\
\hline

 & $^{60}{\rm Zn}(p,\gamma)$ & + & 60 & $5.0\times10^{-5}$ & $2.0\times10^{0}$ \\
\cline{4-6}
 61 &  &  & 120 & $1.1\times10^{-7}$ & $1.1\times10^{-3}$ \\
\cline{2-6}
 & $^{61}{\rm Zn}(p,\gamma)$ & - & 60 & $8.7\times10^{-5}$ & $3.5\times10^{0}$ \\
\cline{4-6}
  &  &  & 120 & $1.4\times10^{-7}$ & $1.4\times10^{-3}$ \\
\hline

 & $^{62}{\rm Zn}(p,\gamma)$ & + & 60 & $1.2\times10^{-4}$ & $4.8\times10^{0}$ \\
\cline{4-6}
 63 &  &  & 120 & $1.3\times10^{-7}$ & $1.3\times10^{-3}$ \\
\cline{2-6}
 & $^{63}{\rm Ga}(p,\gamma)$ & - & 60 & $7.8\times10^{-6}$ & $3.1\times10^{-1}$ \\
\cline{4-6}
  &  &  & 120 & $1.2\times10^{-8}$ & $1.2\times10^{-4}$ \\
\hline

 & $^{64}{\rm Ga}(p,\gamma)$ & + & 60 & $1.7\times10^{-5}$ & $6.8\times10^{-1}$ \\
\cline{4-6}
 65 &  &  & 120 & $2.0\times10^{-8}$ & $2.0\times10^{-4}$ \\
\cline{2-6}
 & $^{65}{\rm Ge}(p,\gamma)$ & - & 60 & $4.0\times10^{-6}$ & $1.6\times10^{-1}$ \\
\cline{4-6}
  &  &  & 120 & $5.2\times10^{-9}$ & $5.2\times10^{-5}$ \\
\hline
\end{tabular}
\end{table}

The evolution of $X(A)$ over the burst reveals two groups. Low-$A$ (29, 31, 33) urca nuclide production primarily occurs in deep burning layers ($y\approx10^{7.5}-10^{8.5}$~g\,cm$^{-2}$) during the rise of the burst light curve ($t\approx0-10$~s), while high-$A$ (55, 57, 59, 61, 63, 65) production primarily occurs in shallower layers ($y\approx10^{6.5}-10^{7.5}$~g\,cm$^{-2}$) during the burst light curve decay. \footnote{The evolution of $X(A)$ for $\log(y)\lesssim6.3$ is complicated by the fact that freshly accreted material will begin to reach these depths during the burst duration, as $y=\dot{M}\Delta t/(4\pi R^{2}$)~\citep{Meis18}. The feature extending to surface around 20\,s is due to convection (see Figure 21 of \citet{Paxt15}).}.
Juxtaposing $X(A)$ for the high-$A$ group with the hydrogen mass fraction $X({\rm H})$ (see Fig.~\ref{fig:XH}), we see that the increase in urca nuclide $X(A)$ at late times ($t\approx60-120$~s) corresponds to a decrease in hydrogen, demonstrating that hydrogen-burning freeze-out\footnote{Defined as burning until proton-capture is limited by the Coulomb barrier at low temperatures.} is responsible for producing these nuclides. The low-$A$ group, on the other hand, is produced in a hotter region where all hydrogen is rapidly consumed. Fig.~\ref{fig:sparks} presents the same evidence from an alternative view, where \ref{fig:sparknocut} contains $X(A)$ averaged over the entire envelope (above the iron substrate in the model) while the average in \ref{fig:spark} is restricted to the shallow region where hydrogen-burning is present at late times.

We note that \citet{Fisk08} present a detailed description of $rp$-process nucleosynthesis for several regions in the neutron star envelope over the evolution of an X-ray burst. By examining their ``above ignition" region, one can see that they also find the synthesis of high-$A$ nuclides in shallow depths at late times in the burst. However, \citet{Fisk08} did not discuss the reaction sequence responsible for high-$A$ production at late times, nor did they discuss the production of odd-$A$ nuclides, as urca cooling was not yet suspected to operate in the accreted neutron star crust. As such, our findings here are not new, they are simply an elaboration on past results.

To understand why the hydrogen-burning freeze-out increases $X(A)$ for the high-$A$ group but not the low-$A$ group, it is necessary to consider proton-capture rates for $rp$-process reactions involved in the production and destruction of these $A$. For a single nucleus, the destruction rate by a particular hydrogen-burning reaction\footnote{Here, $(p,\gamma)$ or $(p,\alpha)$.} in cgs units is
\begin{equation}
    \lambda = \frac{X({\rm H})}{A_{\rm H}}\rho N_{\rm A}\langle\sigma v\rangle,
    \label{eqn:hburnrate}
\end{equation}
where $A_{\rm H}=1$ is the mass number of hydrogen and $N_{\rm A}\langle\sigma v\rangle$ is the reduced reaction rate evaluated at $T$. Table~\ref{tab:lambda} compares $\lambda$ for reactions primarily responsible for producing and destroying the urca nuclides of interest, based on the $rp$-process reaction flow (see Fig.~5 of \citet{Meis19}) and $N_{\rm A}\langle\sigma v\rangle$ from REACLIBv2.2. For demonstration purposes, we use the environment conditions (Figs.~\ref{fig:XH} and~\ref{fig:denstemp}) near the beginning and end of the shallow hydrogen-burning freeze-out, namely corresponding to times $t=60$ and $120$~s, respectively, for $y\approx10^{7}$~g\,cm$^{-2}$. 

For most of the odd-$A$ nuclides in Table~\ref{tab:lambda}, production outpaces destruction. For $A=31,~63$, and 65, this is at least in part because the destruction reaction involves a higher-$Z$ nuclide and therefore a larger Coulomb barrier. For $A=31,~57,~61,$ and 65, the destruction reaction involves transmuting an even-$Z$ odd-$N$ nucleus into an odd-$Z$ odd-$N$ nucleus, which is energetically unfavorable due to nuclear pairing. For $A=29,~33,~55$, and 59, neither of these conditions are fulfilled and so the production rate is less favored relative to destruction, and is even smaller than the destruction rate for $A=33$.

When the production rate exceeds the destruction rate, an enhancement of $X(A)$ is possible. However, when the mean lifetime for an odd-$A$ nucleus $\uptau=1/\lambda_{-}$ is sufficiently short, a substantial fraction of that $X(A)$ will flow into higher-$A$ nuclides. At later times in the burst evolution, $\uptau<1$~s for the low-$A$ group while $\uptau>100$~s for the high-$A$ group. Therefore, hydrogen burning powering the tail of the burst light curve, when $T$ and $X({\rm H})$ are reduced, results in material being transmuted from low-$A$ nuclides toward higher $A$ and subsequently an enhancement of $X(A)$ in the high-$A$ group. This is reminiscent of the late-time formation of the rare-earth peak in $r$-process nucleosynthesis~\citep{Surm97}, except here the Coulomb barrier is responsible for funneling rather than nuclear structure related contours of the nuclear mass surface.

\section{Implications for Calculations of Urca Nuclide Production}
\label{sec:ASTimplications}

Before addressing whether our findings have implications for nuclear physics studies, we first demonstrate that high-$A$ urca nuclide production is sensitive to plausible variations in nuclear reaction rates, and thereby calculation results will be sensitive to input nuclear physics. We focus on example reactions from Table~\ref{tab:lambda} that are potential targets for future nuclear physics experiments, as discussed in Section~\ref{sec:NUCimplications}. Namely, these are $^{55}{\rm Co}(p,\gamma)^{56}{\rm Ni}$, $^{57}{\rm Ni}(p,\gamma)^{58}{\rm Cu}$, $^{62}{\rm Zn}(p,\gamma)^{63}{\rm Ga}$, and $^{64}{\rm Ga}(p,\gamma)^{65}{\rm Ge}$.

\subsection{Results for typical extreme rate variation factors}
\label{ssec:factor100}
In prior works investigating the sensitivity of X-ray burst nucleosynthesis to nuclear reaction rate variations, a factor of 100 rate enhancement or reduction factor has sometimes been adopted, e.g. \citep{Cybu16,Meis17}. The factor of 100 is intentionally generous, but sometimes strains plausibility when doing a realistic assessment of the reaction rate uncertainty. In this spirit, we performed X-ray burst model calculations for the four aforementioned reactions, varying each reaction individually by $\times$100 or $/100$, applying the same factor for the associated reverse reaction. Burst ashes were determined from the envelope composition after a sequence of 20 bursts, as described in \citet{Meis19}. The ratio of $X(A)$ for the two calculations is shown in Fig.~\ref{fig:AshRatios}.

\begin{figure}
    \centering
    \includegraphics[width=1.0\columnwidth]{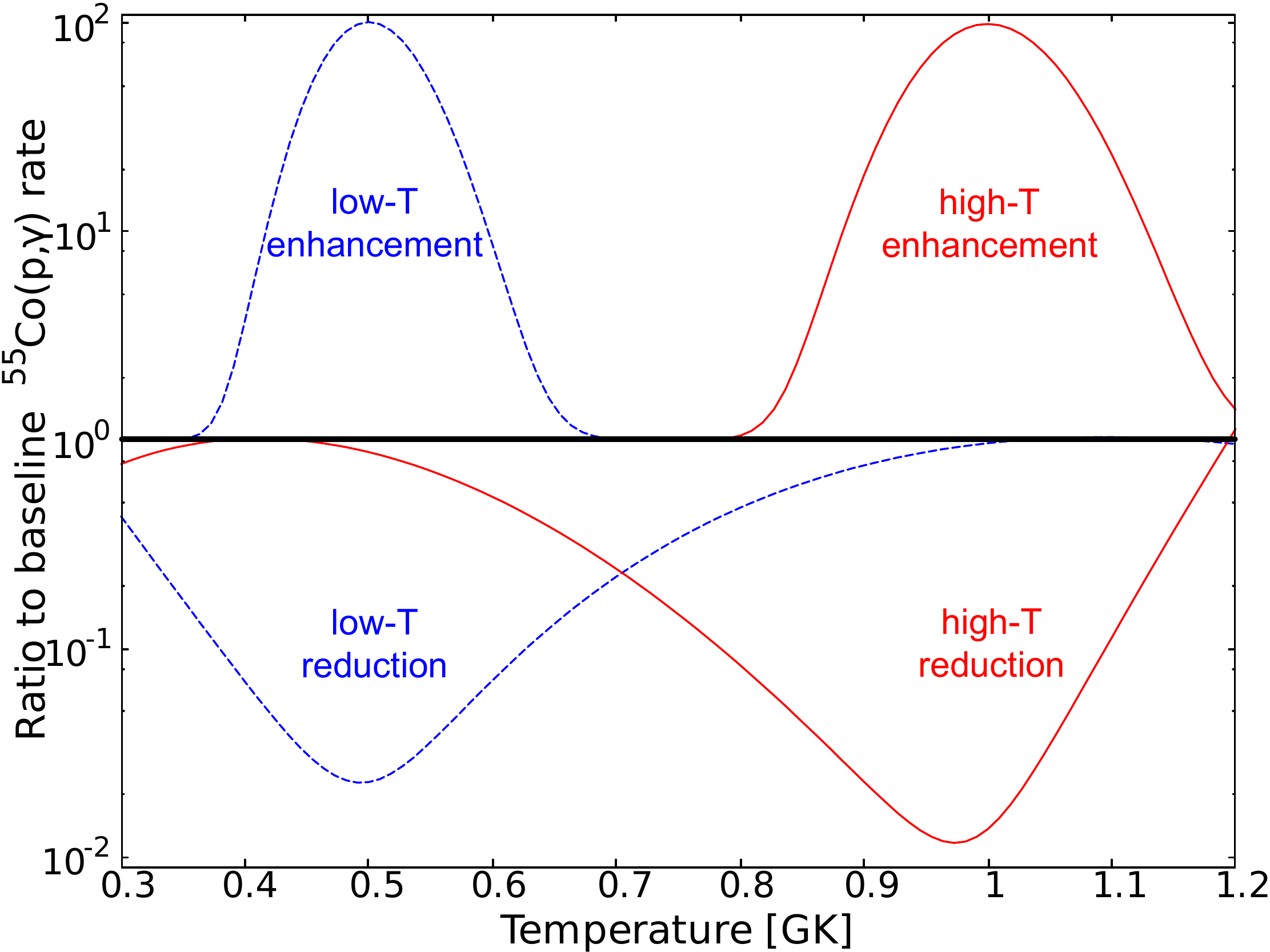}
    \caption{Ratio to the baseline $^{55}{\rm Co}(p,\gamma)$ reaction rate for reaction rates constructed to have a rate enhancement or reduction at low or high temperature. Similarly modified rates were constructed for $^{57}{\rm Ni}(p,\gamma)$, $^{62}{\rm Zn}(p,\gamma)$, and $^{64}{\rm Ga}(p,\gamma)$.}
    \label{fig:RateModifications}
\end{figure}

We also performed calculations with temperature-dependent rate modifications. This isolates the impact of the reaction rate variation at low and high temperatures and demonstrates that high-$A$ urca production is sensitive to the low-temperature reaction rate behavior. We constructed unphysical reaction rates that exhibit a factor of 100 enhancement or reduction near $T\sim0.5$~GK or 1~GK, as shown in Fig.~\ref{fig:RateModifications}. The ratio between burst ashes obtained when using the rate enhancement and reduction at high or low temperature are included in Fig.~\ref{fig:AshRatios}. It is apparent that the low-temperature reaction rate variations drive the change in $X(A)$ in the high-mass region, further demonstrating that high-$A$ urca nuclide production occurs in relatively low-$T$ conditions.

\subsection{An aside on astrophysical reaction rates}

Prior to presenting the more plausible reaction rate variations that we adopted, some discussion is needed connecting the properties of nuclei to nuclear reaction rates. When studying the compound nucleus formed in the fusion of two nuclei in an astrophysical process, the excitation energy region of interest is determined by
$E_{x}=Q+E_{0}\pm\Delta$, where the $Q$-value is the rest mass energy liberated in the reaction and $E_{0}\pm\Delta$ is the Gamow window. The window describes the region in which the convolution of the Maxwell-Boltzmann distribution, describing the nuclear kinetic energy, with the nuclear reaction cross section has the primary contribution from the integrand. A common approximation results in a peak energy
\begin{equation}
    E_{0}=0.122(Z_{1}^{2}Z_{2}^{2}\mu)^{1/3}T_{9}^{2/3}~\rm{MeV},
    \label{eqn:Gamowpeak}
\end{equation}
where $Z_{i}$ are the number of protons in the nuclei participating in the reaction and $\mu=A_{1}A_{2}/(A_{1}+A_{2})$ is their reduced mass, and a $1/e$ width
\begin{equation}
    \Delta=\frac{4}{\sqrt{3}}\sqrt{E_{0}k_{\rm B}T}~\rm{MeV}.
    \label{eqn:Gamowwidth}
\end{equation}
However, for $\sim$GK temperatures, it has been shown that this approximation is too crude and an evaluation of the integrand using a theoretical estimate of the cross section is necessary, often resulting in a window with a lower energy peak than Equation~\ref{eqn:Gamowpeak} and a broader window width than Equation~\ref{eqn:Gamowwidth}. We refer to this energy window as the astrophysical window, using the bounds from ~\citet{Raus10}.

An astrophysical reaction rate per a pair of nuclei $i$ and $j$ is determined from
\begin{equation}
\label{eqn:rxnrate}
    \langle\sigma v\rangle_{ij}=\sqrt{\frac{8}{\pi\mu}}\frac{1}{\left(k_{\rm B}T\right)^{3/2}}\int_{0}^{\infty}\sigma_{ij}(E)E\exp\left(-\frac{E}{k_{\rm B}T}\right),
\end{equation}
where $\sigma_{ij}(E)$ is the reaction cross section between $i$ and $j$ at center of mass energy $E$. If $Q+E\approx E_{x}$ for a state $c$ within the compound nucleus formed by $i+j$, then the reaction rate is characterized by a resonance and the center of mass energy is the resonance energy $E_{\rm r}$. If the state is sufficiently long-lived, so that the energy width obtained by the Heisenberg uncertainty principle is much smaller than $E_{x}$, and well isolated from nearby excited states, then instead the following form can be used
\begin{equation}
\label{eqn:resrate}
 \langle\sigma v\rangle_{ij}=\left(\frac{2\pi}{\mu k_{\rm B}T}\right)^{3/2}\hbar^{2}\omega\gamma_{c}\exp\left(-\frac{E_{\rm r}}{k_{\rm B}T}\right).
\end{equation}
Here $\omega\gamma_{c}$ is the resonance strength evaluated at $E_{\rm r}$. The statistical factor $\omega=(2J_{c}+1)/((2J_{i}+1)(2J_{j}+1))$, referring to the spins of the interacting nuclei and the spin of the state populated in the compound nucleus by $i+j\rightarrow c$. If $i$ is the lower-$A$ nucleus tunneling into $j$ and $k$ is the exit channel (either a low-$A$ nucleus or a photon emitted from $c$),
    $\gamma_{c}=\Gamma_{i}(E)\Gamma_{k}(E)/\Gamma_{\rm tot}(E)$,
where $\Gamma_{i}$ and $\Gamma_{k}$ are the entrance channel and exit channel partial widths, respectively, and $\Gamma_{\rm tot}$ is the sum of all partial widths associated with the decay of $c$.

When multiple resonances are present in the astrophysical window, the reaction rate will be characterized by a sum of the narrow resonance rates, still referred to as a resonant reaction rate. However, when roughly 10 or more resonances are present within the astrophysical window, then a statistical treatment for the reaction is adequate~\citep{Raus97}. This reaction regime is modeled using the Hauser-Feshbach approach, where average resonance properties are used to determine the reaction rate~\citep{Wolf51,Haus52}. In this case, the cross section to use in Equation~\ref{eqn:rxnrate} is  $\sigma_{\rm HF}\propto\lambda^{2}(\langle\mathcal{T}_{i}\rangle\langle\mathcal{T}_{k}\rangle)/\mathcal{T}_{\rm tot}$, where $\lambda$ is the de Broglie wavelength for the entrance channel, $\langle\mathcal{T}_{i}\rangle$ and $\langle\mathcal{T}_{j}\rangle$ are the average transmission coefficients for the entrance and exit channels, respectively, and $\mathcal{T}_{\rm tot}$ is the sum over average transmission coefficients for all channels by which $c$ can decay. Average transmission coefficients are related to average partial widths by $\langle\mathcal{T}\rangle=2\pi\langle\rho\rangle\langle\Gamma\rangle$, where $\langle\rho\rangle$ is the average level density for the nucleus that the reaction channel populates.

\subsection{Results for physically-motivated rate variation factors}
\label{ssec:factorphysical}
In practice, resonant reaction rates often have large uncertainties because they sensitively depend on $\Gamma$ and $E_{r}$ for one or a few states in $c$, which are often poorly constrained. Meanwhile, averaging over the properties of many states present when in the statistical regime tends to wash-out such uncertainties, resulting in an overall lower uncertainty for a statistical nuclear reaction rate.

For physically motivated uncertainties of statistical reaction rates, we adopt the factors from \citet{Raus16} that were based on comparisons to data, where $(p,\gamma)$ reactions are subject to a factor of 2 enhancement and factor of 3 reduction with respect to a baseline Hauser-Feshbach estimate.

For resonant rates, one approach is to determine a Monte Carlo uncertainty by calculating the resonant reaction rate many times while sampling the resonance properties from probability distributions~\citep[e.g.][]{Long12}. While the Monte Carlo approach enables statistical rigor, it can underestimate the uncertainty if missing resonances are not accounted for or if the probability distributions for resonance properties are not sufficiently accurate~\citep{Cava15,SunL2020}.

We instead opt for a simpler approach to estimate a plausible resonant reaction rate enhancement, whereby a single resonance within the astrophysical window is assumed to have the maximum possible $\omega\gamma$, provided by the Wigner limit~\citep{Teic52}. For this case, we assume $\Gamma_{\rm tot}\approx\Gamma_{i}+\Gamma_{k}$ and $\Gamma_{k}>>\Gamma_{i}$, such that $\omega\gamma\approx\omega\Gamma_{i}$. In the Wigner limit,
\begin{equation}
\label{eqn:WigLimit}
    \Gamma_{i,{\rm Wig.}}=\frac{3\hbar^{2}}{2\mu a^{2}}P_{ij}(E,\ell),
\end{equation}
where the channel radius $a$ is often defined as the interaction radius $r_{0}\left(A_{i}^{1/3}+A_{j}^{1/3}\right)$, with $r_{0}$ determined from fits to nuclear radii. Here we use $r_{0}=1.36$~fm, determined from \citet{Ange13}. The penetrability $P_{ij}(E,\ell)$ is the probability of $i$ tunneling through the potential barrier of $j$ with angular momentum transfer $\ell$. The proper calculation of $P_{ij}(E,\ell)$ requires solving the Schr\"{o}dinger equation for scattering of $i$ on the nuclear potential of $j$, but more convenient analytic forms are available that give an order of magnitude estimate when $E$ is below the Coulomb barrier~\citep{Humb87}. Here we use a simple estimate obtained from the WKB approximation for tunneling the Coulomb potential,
\begin{equation}
    \label{eqn:PenCoul}
    P_{ij}(E)=\exp\left(-2\pi\alpha_{\rm fs}Z_{i}Z_{j}\sqrt{\frac{\mu c^{2}}{2E}}\left(1-\frac{4}{\pi}\sqrt{\frac{a}{\frac{Z_{i}Z_{j}e^{2}}{E}}}\right)\right),
\end{equation}
where $\alpha_{\rm fs}$ is the fine structure constant and $e$ is the electron charge.
The centrifugal barrier is included by $P_{ij}(E,\ell)=P_{ij}(E)P_{ij}(\ell)$, where
\begin{equation}
    \label{eqn:PenCent}
    P_{ij}(\ell)=\exp\left(-2\ell(\ell+1)\sqrt{\frac{\hbar^{2}}{2\mu Z_{i}Z_{j}e^{2}a}}\right).
\end{equation}

\begin{figure}
    \centering
     \subfigure[$^{55}{\rm Co}(p,\gamma)^{56}{\rm Ni}$]{\label{fig:co55pg}\includegraphics[width=0.5\textwidth]{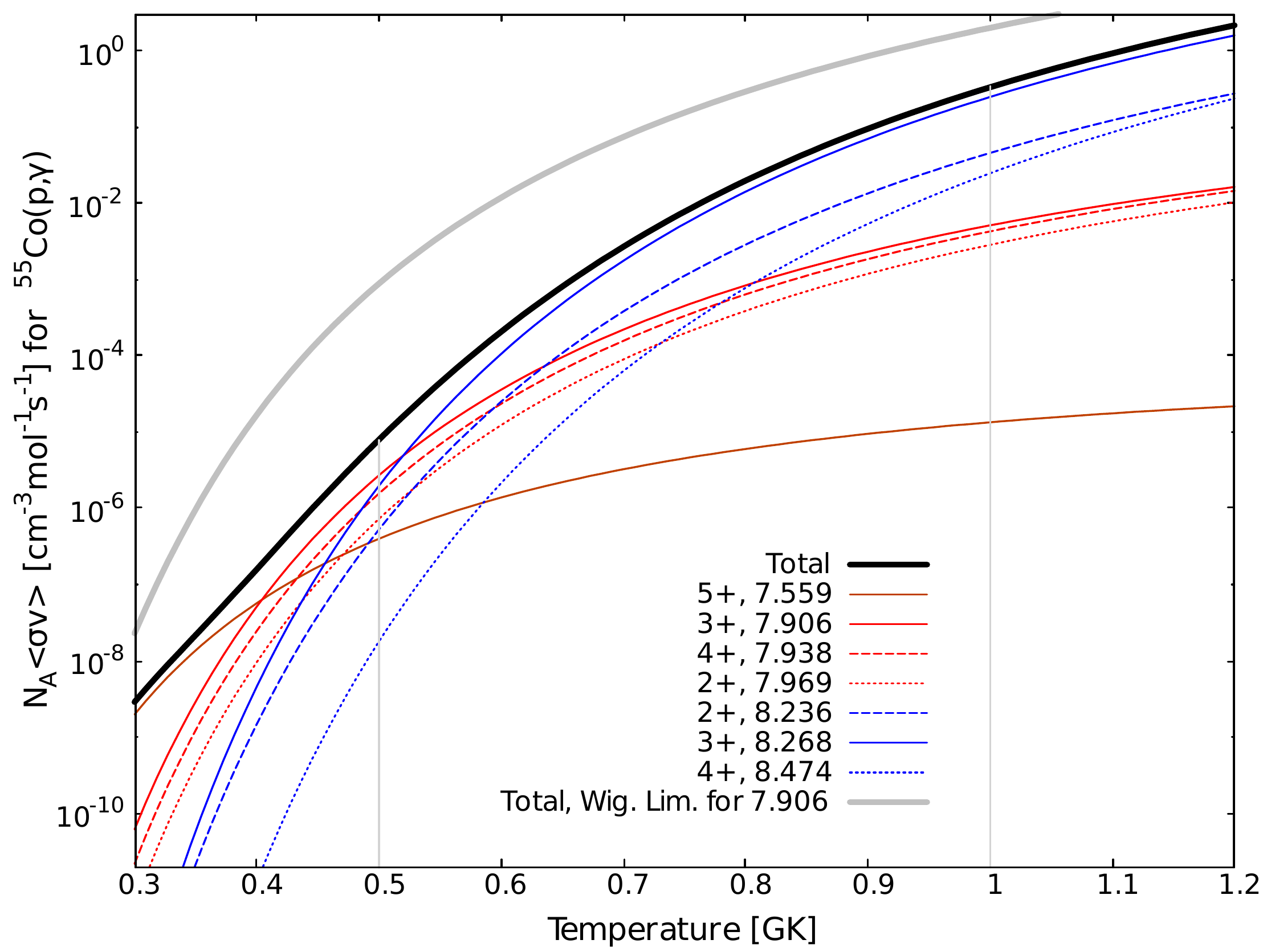}}
  \qquad
  \subfigure[$^{57}{\rm Ni}(p,\gamma)^{58}{\rm Cu}$]{\label{fig:ni57pg}\includegraphics[width=0.5\textwidth]{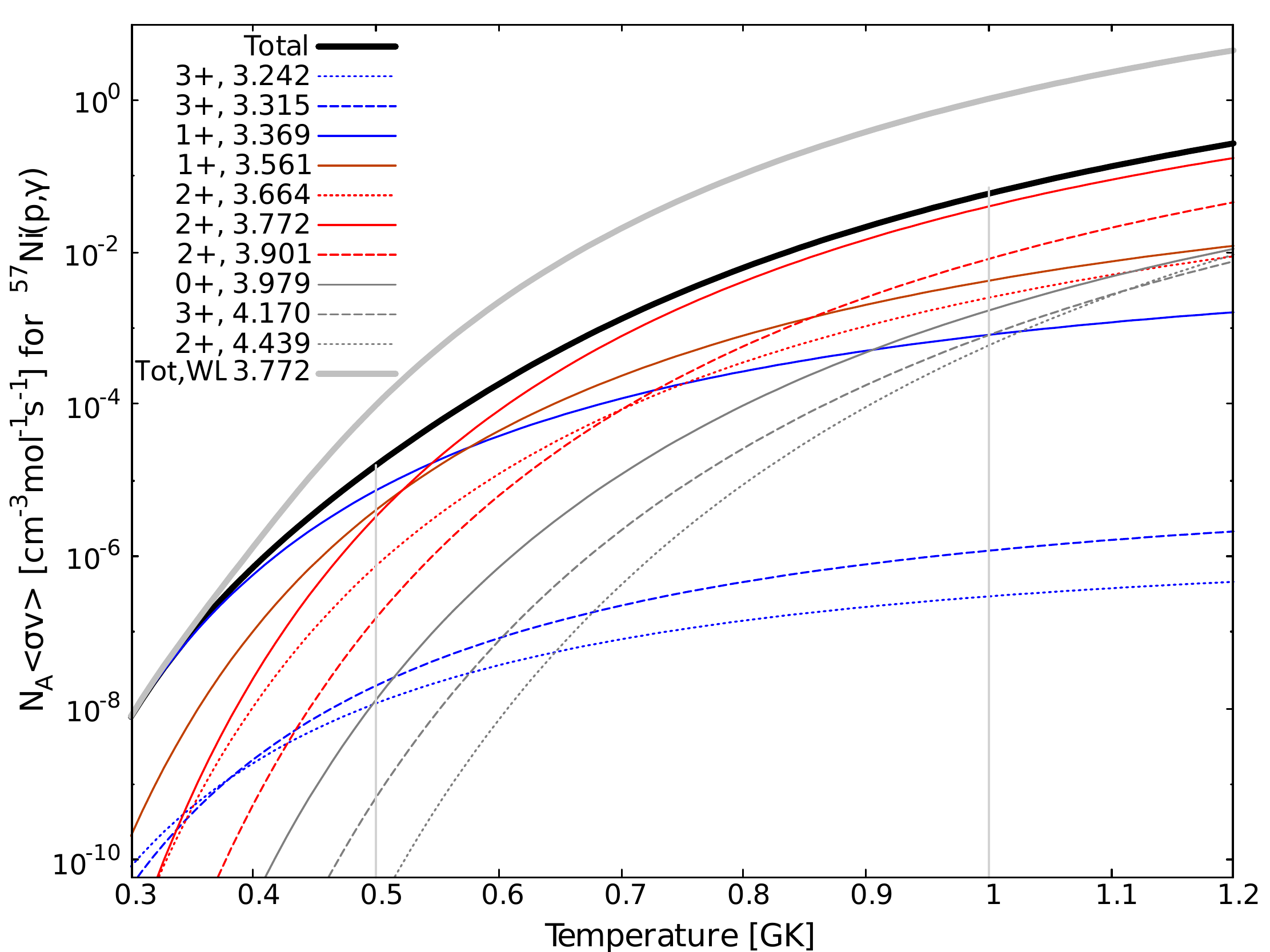}}
    \caption{Astrophysical reaction rates calculated using Eqn.~\ref{eqn:resrate}. The thick solid black line shows the total reaction rate resulting from the sum of the thin lines, which use excited state properties from \citep{Fisk01}, where the legend indicates $E_{x}$ in the compound nucleus. The thick gray line shows the total rate when the indicated excited state has $\omega\gamma$ enhanced to the Wigner limit. Thin vertical gray lines help locate 0.5~GK and 1.0~GK.}
    \label{fig:resrates}
\end{figure}

Spectroscopic constraints are inadequate for the compound nuclides produced in the reactions $^{55}{\rm Co}(p,\gamma)^{56}{\rm Ni}$, $^{57}{\rm Ni}(p,\gamma)^{58}{\rm Cu}$, $^{62}{\rm Zn}(p,\gamma)^{63}{\rm Ga}$, and $^{64}{\rm Ga}(p,\gamma)^{65}{\rm Ge}$. Instead, we use the theoretical results of \citet{Fisk01}, who used shell model calculations to determine excited state properties in the compound nuclides of interest here. According to those calculations, the heuristic for statistical model validity of 10 levels or more per astrophysical window is fulfilled for $^{62}{\rm Zn}(p,\gamma)^{63}{\rm Ga}$ and $^{64}{\rm Ga}(p,\gamma)^{65}{\rm Ge}$, but not $^{55}{\rm Co}(p,\gamma)^{56}{\rm Ni}$ and $^{57}{\rm Ni}(p,\gamma)^{58}{\rm Cu}$. Therefore, we adopt the $\times2$ rate increase and $\times1/3$ rate reduction as plausible uncertainties for these reactions. For $^{55}{\rm Co}(p,\gamma)^{56}{\rm Ni}$ and $^{57}{\rm Ni}(p,\gamma)^{58}{\rm Cu}$, we take a resonance that starts out with a large contribution to the reaction rate  and enhance $\omega\gamma$ to the Wigner limit. The corresponding reaction rates are shown in Fig.~\ref{fig:resrates}, with individual contributions to the total reaction rate, as well as the total rate with the aforementioned resonance enhancement.

\begin{figure*}
    \centering
    \includegraphics[width=1.9\columnwidth]{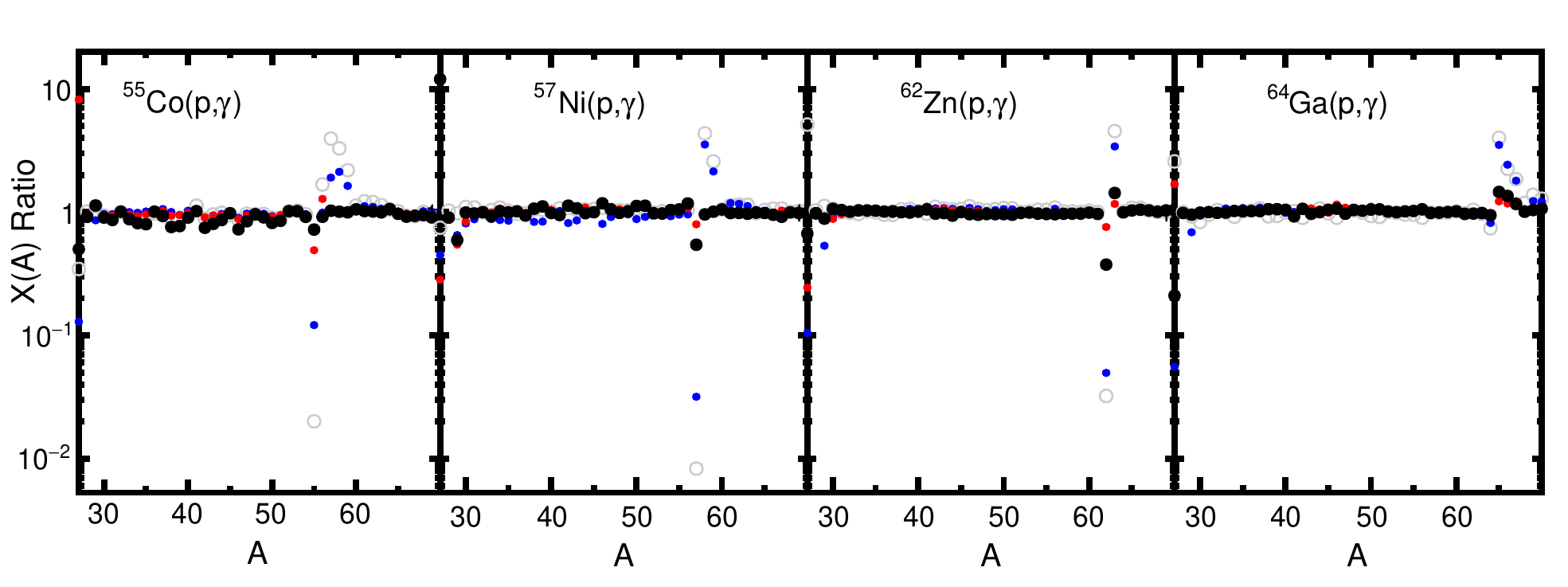}
    \caption{Ratio between $X(A)$ averaged over the envelope after 20 bursts for reaction rate increase/decrease (or increase/baseline),  $\mathcal{X}(A)$, using the rate variation factor 100 (open gray circles), 100 at low temperature (small blue circles), 100 at high temperature (small red circles), and physically motivated (black filled circles). For $^{55}{\rm Co}(p,\gamma)$ and $^{57}{\rm Ni}(p,\gamma)$, the $X(A)$ ratio with the physically-motivated rate variation factor is increase/baseline, for all others it is increase/decrease. See Sections~\ref{ssec:factor100} and \ref{ssec:factorphysical} for a description of the temperature-dependent and physically-motivated reaction rate variation factors, respectively.}
    \label{fig:AshRatios}
\end{figure*}

The impact on $X(A)$ for these physically motivated reaction rate uncertainties is shown in Fig.~\ref{fig:AshRatios}. Here we consider the ratio of $X(A)$ for the reaction rate enhancement to $X(A)$ for the reaction rate reduction (or baseline, in the case of the physically-motivated reaction rate uncertainty for $^{55}{\rm Co}(p,\gamma)$ and $^{57}{\rm Ni}(p,\gamma)$\footnote{For resonant reactions, the physically-motivated reaction rate enhancement is more straightforward to estimate than a rate decrease. This is because a rate having one of several resonances near the Wigner limit is within reason, whereas a significant reaction rate decrease would require all relevant resonances to be well below the average expected width. More rigorous upper and lower bounds would require a more sophisticated approach, such as the High Fidelity Resonance approach~\citep{Roch17}, but applied to proton-capture. We leave such work to a future sensitivity study.}), which we will denote as $\mathcal{X}(A)\equiv \frac{X_{\rm enhancement}(A)}{X_{\rm decrease/baseline}(A)}$. The impact of varying $^{55}{\rm Co}(p,\gamma)$ is relatively modest for the physically motivated rate variation, where $\mathcal{X}(55)=0.7$, as compared to $\mathcal{X}(55)=0.02$ for a reaction rate variation factor of 100. The rate variation factor of 100 also leads to $\mathcal{X}(57)=3.8$ and $\mathcal{X}(59)=2.1$. The physically motivated rate variation for $^{57}{\rm Ni}(p,\gamma)$ results in $\mathcal{X}(57)=0.5$, whereas $\mathcal{X}(57)=0.01$ for the rate variation factor of 100. The rate variation factor of 100 also leads to a $\mathcal{X}(59)=2.4$. For $^{62}{\rm Zn}(p,\gamma)$, $\mathcal{X}(63)=1.3$ for the physically-motivated reaction rate variation, whereas a factor of 100 rate variation leads to a $\mathcal{X}(63)=4.4$. For $^{64}{\rm Ga}(p,\gamma)$, the physically motivated rate enhancement results in $\mathcal{X}(65)=1.4$ and $\mathcal{X}(67)=1.2$. For the usual factor of 100 rate variation factor, $\mathcal{X}$ is 3.8, 1.8, and 1.3 for $A=65$, 67, and 69, respectively.

\subsection{Are these \texorpdfstring{$X(A)$}{X(A)} impacts significant?}

Prior works have demonstrated some of the astrophysical consequences of urca cooling in the accreted neutron star crust for realistic estimates of $L_{\nu}$. \citet{Deib16} showed that urca cooling can alter the carbon-ignition depth for X-ray superbursts, widening the discrepancy between the ignition depth inferred from observations and model calculations. \citet{Meis17} showed that urca cooling could lead to observable features in the light curve of a neutron star cooling after sustained accretion ceases.

Whether the impact of a nuclear reaction rate uncertainty on $X(A)$ results in a significant uncertainty contribution to $L_{\nu}$ depends on other influences on $X(A)$ as well as the relative uncertainty contribution from other inputs in Equation~\ref{eqn:Lnu}. 

$X(A)$ is also sensitive to changes in assumed astrophysical conditions for a given bursting source. Recent works have constrained the astrophysical conditions for the source GS 1826-24 by performing model-observation comparisons of the X-ray burst light curve over several bursting epochs~\citep{Meis18b,John20}. Though performed with different codes, the \citet{Meis18b} ({\tt MESA}) and \citet{John20} ({\tt KEPLER}) results are largely consistent. For instance, the uncertainty in the accretion rate considering the uncertainties in each individual work, along with the discrepancy between the two results, is roughly 10\%. For a similar change in accretion rate, \citet{Meis19} found $X(A)$ for urca nuclides is changed roughly 20\% or less. According to the \citet{Meis18b} calculations, using the upper limit for $X_{\rm H}$ from \citet{John20} would result in modifications of $X(A)$ for urca nuclides of roughly 30-50\%. We caution that a rigorous benchmarking of the {\tt MESA} and {\tt KEPLER} codes has not been performed to date. Thus comparing the codes' results provides only a rough guide for the uncertainties that can potentially be achieved in X-ray burst light curve model-observation comparisons.

For a fixed set of astrophysical conditions, these enter through $L_{34}(Z,A)$: 
\begin{equation}
\label{eqn:L34}
L_{34}(Z,A)\approx 0.87\left(\frac{10^{6}~{\rm{s}}}{ft}\right)\left(\frac{56}{A}\right)\left(\frac{Q_{\rm{EC}}(Z,A)}{4~{\rm{MeV}}}\right)^{5}\left(\frac{\langle F\rangle^{*}}{0.5}\right), \\
\end{equation}
where $Q_{\rm EC}(Z,A)$ is the $e^{-}$-capture $Q$-value for $(Z,A)\rightarrow(Z-1,A)$, $\langle F\rangle^{*}\equiv\langle F\rangle^{+}\langle
F\rangle^{-}/(\langle F\rangle^{+}+\langle F\rangle^{-})$, and $\langle F\rangle^{\pm}\approx2\pi\alpha_f
Z/|1-\exp(\mp2\pi\alpha_f Z)|$ is the Coulomb factor. The comparative half-life $ft$ should be implemented as $ft=({ft_{e^{-}\rm{-capture}}+ft_{\beta^{-}}})/{2}$, since the degeneracy of the parent state impacts the transition rate.

The uncertain quantities in Equation~\ref{eqn:L34} are $Q_{\rm EC}$ and $ft$. The majority of nuclear masses required to calculate $Q_{\rm EC}$ for the relevant nuclei are already well-constrained~\citep{Meis20a}, with most contributing on the order of 10\% uncertainty to $L_{\nu}$. It is anticipated that in the near future higher precision constraints will further reduce these mass uncertainties such that the uncertainty contribution of $Q_{\rm EC}$ to $L_{\nu}$ will be less than one percent \citep{Meis20b}. Experimental $ft$ constraints are presently more sparse. However, with the significant increase in statistics at near-future radioactive ion beam facilities, it should be possible to apply current techniques~\citep[e.g.][]{Baum87} to obtain uncertainties on the tens of percent level.

Therefore, any nuclear reaction rate variations that alter $X(A)$ of urca nuclides, and thereby $L_{\nu}$, by tens of percent or more can be considered significant. In the context of the previously discussed impact on $X(A)$ from nuclear reaction rate variations, this would be $\mathcal{X}(A)\lesssim0.8$ or $\mathcal{X}(A)\gtrsim1.2$. By this criterion, all of the reaction rate variations described in this section for $^{55}{\rm Co}(p,\gamma)^{56}{\rm Ni}$, $^{57}{\rm Ni}(p,\gamma)^{58}{\rm Cu}$, $^{62}{\rm Zn}(p,\gamma)^{63}{\rm Ga}$, and $^{64}{\rm Ga}(p,\gamma)^{65}{\rm Ge}$ are significant. We stress that more detailed spectroscopy is needed for $^{56}{\rm Ni}$, $^{58}{\rm Cu}$, $^{63}{\rm Ga}$, and $^{65}{\rm Ge}$ so that more rigorous reaction rate uncertainty assessments can be performed.

\subsection{Are these the most significant \texorpdfstring{$X(A)$}{X(A)} impacts?}

Likely not. The reactions we investigate in this work are based on the nuclear reaction network flow for the $rp$-process. However, a full reaction rate sensitivity study, varying a large number of reaction rates within estimated uncertainties and quantifying the impact on $X(A)$, would be required to produced a ranked list of reaction rate importance. We leave that significant undertaking for a future work. The fact that we find important sensitivities in the present work indicates that such a sensitivity study would be worthwhile.

\subsection{Pertinence for sensitivity study results}

The different temperature range of interest for urca nuclide production, as opposed to temperatures of interest for X-ray burst light curve impacts, explains why certain reaction rates impact model calculation results for the light curve but not the burst ashes. For instance, the reaction rate sensitivity studies of \citet{Cybu16} and \citet{Meis19} each found that the competition between the $^{59}{\rm Cu}(p,\gamma)$ and $^{59}{\rm Cu}(p,\alpha)$ reaction rates plays a key role in the decaying phase of the X-ray burst light curve. However, neither study observed an impact on $X(59)$ from changes to the $^{59}{\rm Cu}(p,\alpha)$ reaction rate\footnote{The post-processing study of \citet{Pari08} had similar findings, where $^{59}{\rm Cu}(p,\alpha)$ impacted nuclear energy generation but not $X(59)$, except for one trajectory using relatively high temperatures and very high metallicity. The high-metallicity conditions lead to exceptionally hydrogen-rich bursts, which will in general enhance nuclear burning of high-$A$ nuclides~\citep{Jose10,Meis19}.}.

\begin{figure}
    \centering
    \includegraphics[width=\columnwidth]{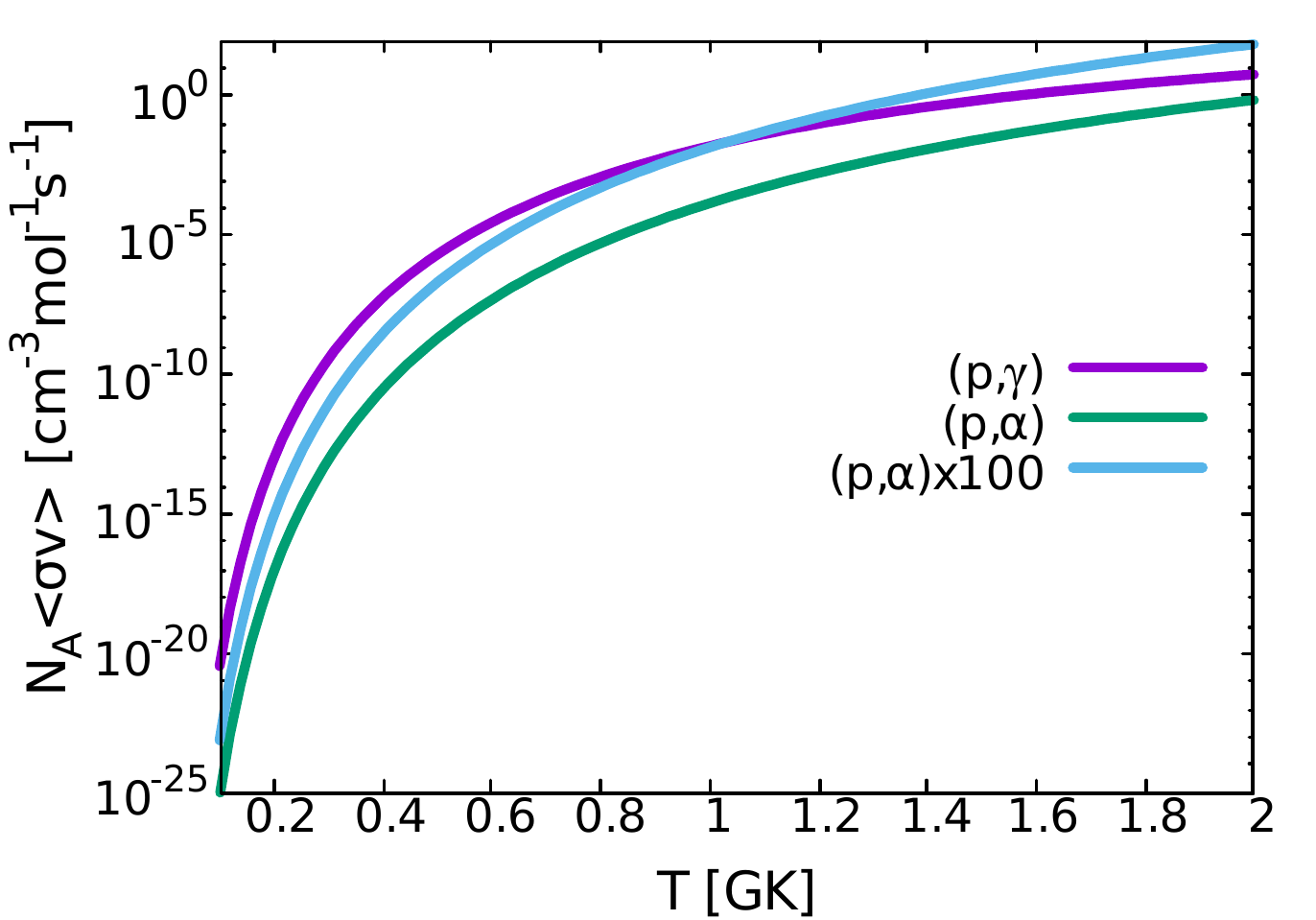}
    \caption{Reaction rates for $^{59}{\rm Cu}+p$ from REACLIBv2.2 label {\tt ths8}~\citep{Cybu10}, acquired from \url{https://reaclib.jinaweb.org/}.}
    \label{fig:cu59rates}
\end{figure}

Fig.~\ref{fig:cu59rates} shows that when the $^{59}{\rm Cu}(p,\alpha)$ reaction rate is enhanced by a 
factor of 100 from the nominal best estimate, as performed by \citet{Meis19} and \citet{Cybu16},
then the $(p,\gamma)$ and $(p,\alpha)$ rates are comparable for $T\approx1$~GK. At that temperature, of relevance for light curve impacts, the two reactions would compete in terms of reaction flow. However, for the $T\lesssim0.6$~GK conditions of relevance for urca nuclide production, the enhanced $(p,\alpha)$ rate is an order of magnitude or more lower than the $(p,\gamma)$ rate, meaning a minor fraction of the overall reaction flow will proceed through this path. Hence, in this example the details of $(p,\alpha)$ versus $(p,\gamma)$ rate competition are important for understanding light curve impacts, but not $A=59$ urca nuclide production.

Note that the lack of influence on urca nuclide production does not preclude the influence of a reaction rate on overall nucleosynthesis. A reaction like $^{59}{\rm Cu}(p,\alpha)$ that influences nuclear energy generation, and therefore the X-ray burst light curve, will modify the thermodynamic conditions of the burst environment. The modified conditions, e.g. maintaining a higher temperature for longer time, will subsequently modify the $rp$-process path, leading to changes in $X(A)$ far from $A$ of the heavy nuclide participating in the influential reaction~\citep{Pari08,Cybu16,Meis19}. 

\section{Implications for Nuclear Physics Studies}
\label{sec:NUCimplications}

An important consequence of the previous section is that high-$A$ urca nuclides are primarily produced where $T\lesssim0.6$~GK. This is much lower temperature than is typically associated with impacts of reactions involving high-$A$ nuclides on the X-ray burst light curve. For instance, Fig.~1 of \citet{Meis19} shows that $^{59}{\rm Cu}(p,\gamma)^{60}{\rm Zn}$ alters the light curve by stalling or enhancing nuclear burning around $t\approx10$~s after burst ignition. Comparing to Figs.~\ref{fig:a59} and \ref{fig:temp}, we see that this nuclear burning occurs at $y\approx10^{8}$~g\,cm$^{-2}$, where $T\approx1.0$~GK. 

Since lower temperatures are of interest for studies constraining high-$A$ urca nuclide production relative to studies constraining light curve impacts, experimental designs may need to be modified relative to studies focused on light curve impacts. We stress that for many cases our findings will not change experimental designs. Below we detail a few examples where our findings may have an impact, to demonstrate that the temperature difference between high-$A$ urca nuclide production and X-ray burst light curve impacts is worth keeping in mind during the planning phase of nuclear physics experiments. These examples require some speculation, as spectroscopy for the relevant nuclides is currently insufficient and some of the experimental devices intended to perform the described experiments are still under development. 

For the reactions of relevance here, direct measurements will require inverse kinematics, where a radioactive ion beam impinges on a hydrogen target. Radiative capture reactions on short-lived nuclides tend to have low yields and high $\gamma$-backgrounds, which favors the use of recoil separators~\citep{Ruiz14}. For the purposes of our estimates below, we will consider hypothetical future measurements using the JENSA gas-jet target~\citep{Schm18} coupled to the SECAR recoil separator at the upcoming Facility for Rare Isotope Beams (FRIB)~\citep{Berg18}, as this will likely be the first set-up where direct measurements of $rp$-process reactions in the astrophysical energy window will be possible. However, we provide all relevant experimental details such that our estimates could be applied to other facilities as they come online in the future, e.g. CRYRING at the Facility for Antiproton and Ion Research (FAIR)~\citep{Lest16}. We anticipate that the length of a typical SECAR experiment at FRIB will be limited to 10~days of beam-on-target based on the precedent established at current radioactive ion beam facilities.

\subsection{Implications for resonance strength measurements}

When an astrophysical reaction rate is well described by individual resonance contributions, direct measurements aim to constrain $\omega\gamma$ (and often $E_{\rm r}$) of the relevant resonances. A large number of beam nuclei $N_{\rm b}$ is impinged on a target yielding $Y_{\rm res}$ reactions per beam particle. When the energy loss of the beam in the target is much greater than the resonance width, the inverse kinematics thick target yield formula can be used:
\begin{equation}
    Y_{\rm res}=\frac{\omega\gamma A_{\rm b}\lambda^{2}}{2\epsilon\mu},
\end{equation}
where $A_{\rm b}$ is the mass number of the beam nucleus, $\epsilon$ is the laboratory-frame stopping power, and both $\omega\gamma$ and $\lambda$ are in the center-of-mass frame. For all the cases considered below, we use $\epsilon\approx2\times10^{-13}$~eV\,cm$^{2}$, based on the {\tt ATIMA} results as calculated by the {\tt LISE++} code~\citep{Tara16}.

The number of detected reaction events $N_{\rm evt}$ will be the product of the incident beam intensity $I_{\rm b}=N_{\rm b}/t_{\rm meas}$, with measurement time $t_{\rm meas}$, and $Y_{\rm res}$, correcting for the fraction of recoil nuclides in the charge-state for which the recoil separator is tuned $\eta_{\rm cs}$, recoil transmission $\eta_{\rm t}$, recoil detection efficiency $\eta_{\rm det}$, and coincident $\gamma$-detection efficiency $\eta_{\gamma}$. Therefore,
\begin{equation}
    N_{\rm evt}=I_{\rm b}t_{\rm meas}Y_{\rm res}\eta,
    \label{eqn:Nevt}
\end{equation}
where $\eta=\eta_{\rm cs}\eta_{\rm t}\eta_{\rm det}\eta_{\gamma}$ and, for the cases under consideration below, we estimate $\eta_{\rm cs}=0.25$~\citep{Shim92}, $\eta_{\rm t}=1$, $\eta_{\rm det}=1$, and $\eta_{\gamma}=0.5$~\citep{Bish03,Chri13,Chri18}.

Fig.~\ref{fig:co55pg} shows that, when assuming resonance properties of \citet{Fisk01}, the $^{55}{\rm Co}(p,\gamma)^{56}{\rm Ni}$ reaction rate has the most significant contributions from populating the $^{56}{\rm Ni}$ excited states at $E_{x}=$7.906, 7.938, and 8.268~MeV at 0.5~GK, whereas at 1.0GK, the state at 8.268~MeV provides the dominant contribution. For the anticipated FRIB reaccelerated beam intensity\footnote{FRIB rate calculator v2.01 : \url{https://groups.nscl.msu.edu/frib/rates/fribrates.html}.} of $^{55}{\rm Co}$, $I_{\rm b}=1.4\times10^{9}~{\rm s}^{-1}$, the yield from the 8.268~MeV state could be measured to 10\% statistics in tens of minutes. However, considerably more time would be required to measure $\omega\gamma$ from the two lower $E_{x}$ contributions\footnote{The gas-jet target density would have to be low enough such that energy loss would not preclude distinguishing the yield from the two states.}. For the \citet{Fisk01} $\omega\gamma$, it would take the full hypothetical 240~hour $t_{\rm meas}$ in order to achieve 20\% precision for both resonances. Hence, based on these theoretical resonance properties and as-yet-unproven radioactive ion beam facility capabilities, the requested measurement time would be vastly different depending on if one is interested in the impacts of $^{55}{\rm Co}(p,\gamma)$ on X-ray burst light curves or high-$A$ urca $X(A)$.

Fig.~\ref{fig:ni57pg} highlights the case of $^{57}{\rm Ni}(p,\gamma)^{58}{\rm Cu}$, where the reaction rate at 1~GK primarily comes from the contribution of the hypothetical 3.772 and 3.901~MeV states in $^{58}{\rm Cu}$, while at 0.5~GK there most significant contributions come from $E_{x}=$3.369, 3.561, and 3.772~MeV. For the predicted $I_{\rm b}=1.4\times10^{9}~{\rm s}^{-1}$ and \citet{Fisk01} $\omega\gamma$, 30~hours of on-target time would be required to measure $Y_{\rm res}$ for the 3.772 and 3.901~MeV states to 10\% statistical precision. Alternatively, if the reaction rate at 0.5~GK is of primary interest, then measuring the 3.561~MeV state would require an additional 210~hours to obtain $\sim$$20-500$ events from this resonance, assuming $\omega\gamma$ falls between the \citet{Fisk01} calculation result and the Wigner limit.

Of course, for many cases measurement plans would not be affected by considering an environment temperature of 0.5~GK instead of 1.0~GK. An example is $^{54}{\rm Co}(p,\gamma)^{55}{\rm Ni}$, whose reaction rate across this entire temperature range is primarily determined by the resonant reaction into the 5.273~MeV state of $^{55}{\rm Ni}$, when using the nuclear structure estimates of \citet{Fisk01}.

\subsection{Implications for cross section measurements}

When the nuclear level density of the compound nucleus is sufficiently high, making the statistical approximation valid, direct measurements aim to determine the nuclear reaction cross section, $\sigma$. When $\sigma$ and $\epsilon$ are expected to be relatively constant over the reaction energy spanned by energy loss in the target, the yield is described by
\begin{equation}
    Y_{\rm thin}=n_{\rm t}\sigma,
\end{equation}
where $n_{\rm t}=10^{19}$~atoms\,cm$^{-2}$ is the areal density assumed for the JENSA hydrogen gas-jet target. To determine $N_{\rm evt}$, Equation~\ref{eqn:Nevt} applies, substituting $Y_{\rm thin}$ in place of $Y_{\rm res}$.

\begin{figure}
    \centering
    \includegraphics[width=\columnwidth]{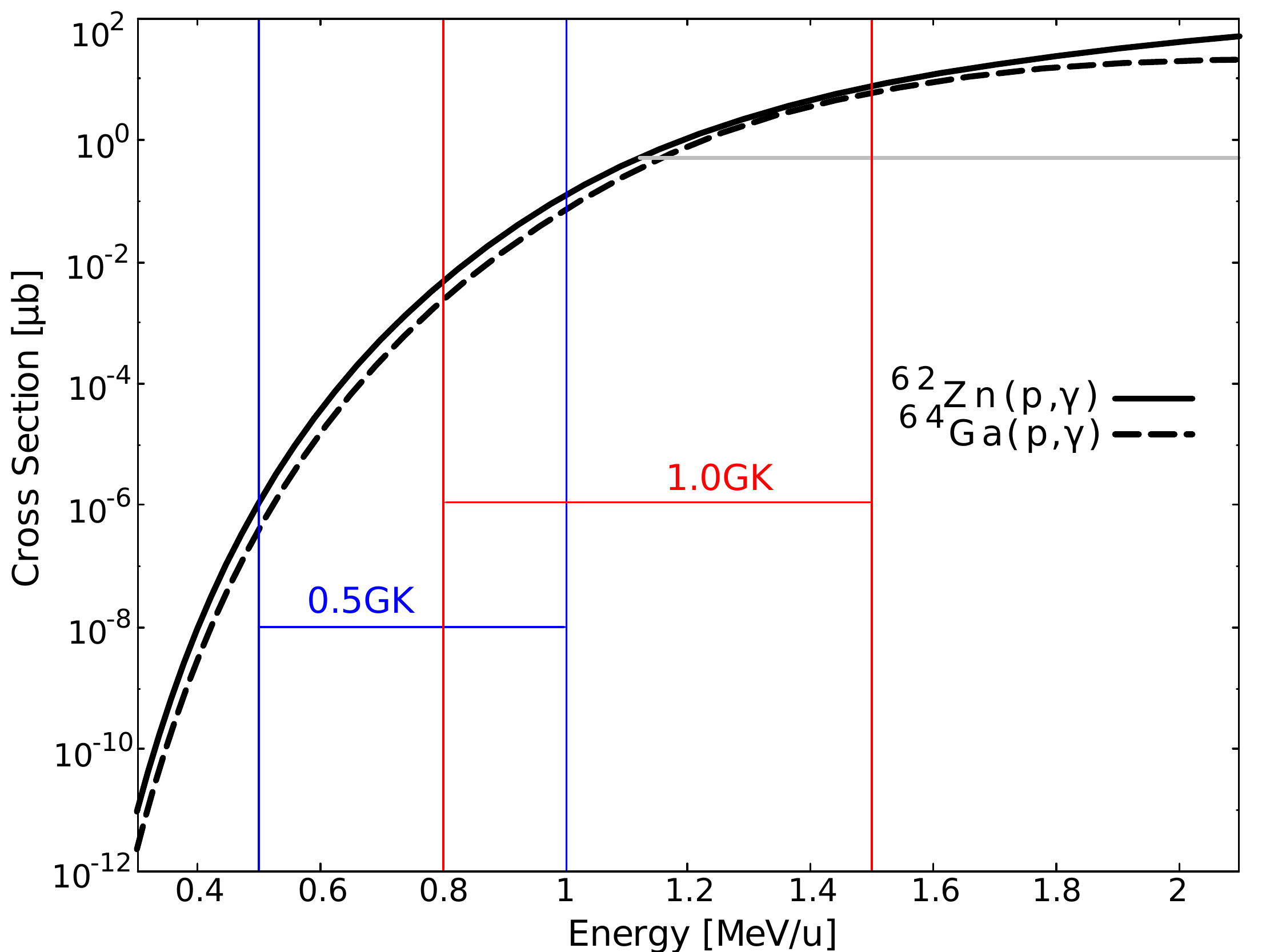}
    \caption{Hauser-Feshbach cross sections for $^{62}{\rm Zn}(p,\gamma)$ (solid) and $^{64}{\rm Ga}(p,\gamma)$ (dashed) from \citet{Raus01} with the 0.5~GK and 1.0~GK astrophysical windows indicated. The gray horizontal line indicates the energy range in which $N_{\rm evt}$ could be measured to 10\% statistical precision in one day or less.}
    \label{fig:CSstat}
\end{figure}

Fig.~\ref{fig:CSstat} focuses on $^{62}{\rm Zn}(p,\gamma)^{63}{\rm Ga}$ and $^{64}{\rm Ga}(p,\gamma)^{65}{\rm Ge}$, using the Hauser-Feshbach $\sigma$ calculated by \citet{Raus01} and $I_{\rm b}=1.4\times10^{9}~{\rm s}^{-1}$. For both cases, 10\% statistical precision in $Y_{\rm thin}$ could be achieved within a day for a measurement energy reaching roughly half-way into the 1.0~GK astrophysical window, whereas nearly the entire plausible measurement time would be required to make a single measurement near the upper end of the 0.5~GK window.

While measuring at the higher energies may help constrain $\sigma$ at lower energies, the constraints may not be particularly stringent. As discussed in the following subsection, the sensitivity of Hauser-Feshbach results to various input physics changes substantially as a function of energy for $\sigma_{\rm HF}$ and temperature for $N_{\rm A}\langle\sigma v\rangle$. For $^{62}{\rm Zn}(p,\gamma)$ and $^{64}{\rm Ga}(p,\gamma)$, $\sigma_{\rm HF}$ in the measurable region is significantly impacted by uncertainties in the $\gamma$-strength function, whereas in the 0.5~GK astrophysical window, $\sigma_{\rm HF}$ is essentially only sensitive to the proton optical model potential~\citep{Raus12}.

\subsection{Implications for indirect measurements}
Often times direct measurements of nuclear reaction data in the astrophysical window are not possible due to insufficient yields, in which case indirect techniques must be applied. For statistical reaction rates, this consists of constraining the quantities required to calculate $\mathcal{T}$, in particular the nuclear level density, optical model potential for particle channels, and the $\gamma$-strength function for photon emission. Generally, indirect experimental approaches are well suited to constraining one or two of these quantities, but seldom all three. For example, elastic scattering can be used to constrain an optical model potential~\citep[e.g.][]{Patr12}, whereas the $\beta$-Oslo technique simultaneously constrains a nucleus's level density and $\gamma$-strength function~\citep{Spyr14}.

The sensitivity of $\sigma_{\rm HF}$ to each of these inputs is energy dependent~\citep{Raus12}. For a $(p,\gamma)$ reaction near the reaction threshold, the Coulomb barrier is substantial and so $P_{ij}(E)$ is small. As such, $\langle\Gamma_{p}\rangle<<\langle\Gamma_{\gamma}\rangle$ and so $\mathcal{T}_{\rm tot}\approx\langle\mathcal{T}_{p}\rangle+\langle\mathcal{T}_{\gamma}\rangle\approx\langle\mathcal{T}_{\gamma}\rangle$. This means $\sigma_{\rm HF}\propto\lambda^{2}\langle\mathcal{T}_{p}\rangle$  at low energies, and consequently, at low temperatures the $(p,\gamma)$ reaction rate is mostly sensitive to the proton optical model potential. At higher temperatures the Coulomb barrier is no longer a significant impediment and the $\gamma$-strength function, which determines $\langle\mathcal{T}_{\gamma}\rangle$, is the most important input to the statistical reaction rate calculation.

Consider the $^{61}{\rm Zn}(p,\gamma)^{62}{\rm Ga}$ reaction rate. \citet{Raus12} found that this rate is almost exclusively sensitive to the proton optical model potential at 0.5~GK, whereas at 1.0~GK there is roughly equal sensitivity to the $\gamma$-strength function. \citet{Cybu16} found that $^{61}{\rm Zn}(p,\gamma)$ does not significantly impact the X-ray burst light curve, but does significantly impact $X(61)$ and $X(63)$ in the burst ashes. Having shown that high-$A$ urca nuclide production occurs near 0.5~GK, we conclude that a measurement of the $^{62}{\rm Ga}$ $\gamma$-strength function is not necessary for X-ray burst applications, but instead future measurements should work to constrain the $^{61}{\rm Zn}+p$ proton optical model potential.

\section{Conclusion}
\label{sec:concl}
In summary, we investigated the conditions responsible for urca nuclide production in Type-I X-ray bursts using model calculations performed with the code {\tt MESA}. Using conditions that reproduce observables from the source GS 1826-24, we follow the nucleosynthesis of a single X-ray burst at the end of a sequence of twenty. Our results show that urca nuclides with $A=55, 57, 59, 61, 63,$ and 65 are produced at late times in the evolution of the X-ray burst light curve in environment temperatures near 0.4--0.6~GK. This is cooler than the roughly 1~GK conditions typically relevant for X-ray burst light curve impacts from nuclear reactions involving high-$A$ nuclides. Our results also show that high-$A$ urca nuclide production is sensitive to plausible variations in nuclear reaction rates, with impacts that could alter model-observation comparisons.

These results could impact plans for nuclear physics studies aimed at reducing the uncertainty in the production of urca nuclides during X-ray bursts. The relatively lower temperature range of interest can shift priorities of direct resonance strength measurements to focus on lower-lying excitation energy regions in the compound nuclei than would be of interest for studies motivated by impacts on the X-ray burst light curve. Similarly, direct cross section measurements may require much more measurement time to be within the astrophysical window. Or, such measurements may be altogether unfeasible. Plans for indirect measurements may also have to be modified, as statistical reaction rate sensitivities differ between low and high temperature conditions. These results also explain why some nuclear reaction rates significantly impact the burst light curve, but not $X(A)$ for the $A$ of the heavy nuclide participating in the reaction. 

Ultimately, our findings will help in planning nuclear physics efforts aimed at better constraining urca nuclide production in X-ray bursts and thereby improve constraints on the thermal properties of the accreted neutron star crust.

\section*{Acknowledgements}

We thank the developers of the {\tt MESA} software and contributors to the
{\tt MESA} Marketplace
({\tt http://cococubed.asu.edu/mesa\_market/}) and the associated
forum, who enabled the
calculations presented here.
This work was supported by the U.S. Department of Energy under grants DE-FG02-88ER40387 and DESC0019042.
We benefited from support by the National
Science Foundation under grant PHY-1430152 (Joint Institute
for Nuclear Astrophysics--Center for the Evolution
of the Elements).

\section*{Data Availability}

{\tt MESA} data were generated using input files available at https://doi.org/10.5281/zenodo.2598715. The data are available upon request.



\bibliographystyle{mnras}
\bibliography{References}








\bsp	
\label{lastpage}
\end{document}